\documentclass[preprint,12pt]{elsarticle}
\journal{Computer Communications}

\usepackage{times}
\usepackage{amssymb}
\usepackage{pifont}

\usepackage{amsthm}
\usepackage{amsmath}
\usepackage{graphicx}
\usepackage{epstopdf}
\usepackage{caption}
\DeclareCaptionType{copyrightbox}
\usepackage{subcaption}
\usepackage[ruled, vlined]{algorithm2e}
\usepackage{array}
\usepackage{enumerate}
\usepackage{comment}
\usepackage{multicol}
\usepackage{multirow}
\usepackage{balance}
\usepackage[table]{xcolor}
\usepackage{url}
\usepackage{verbatim}
\usepackage{microtype}
\journal{Ad Hoc Networks}

\newcommand{\eat}[1]{}

\newcommand{\algname}[1]{\textsf{\footnotesize #1}}

{
\theoremstyle{definition}

\newtheorem{example}{Example}

}

\newcommand{\tobj}[1]{\textsf{\tiny #1}}

\begin{document}

\begin{frontmatter}

\title{A General Model for MAC Protocol Selection\\in Wireless Sensor Networks}

\author{Abolfazl Asudeh\corref{cor1} \fnref{label1}}
\ead{ab.asudeh@mavs.uta.edu}
\author{Gergely V. Zaruba \fnref{label1}}
\ead{zaruba@uta.edu}
\author{Sajal K. Das \fnref{label2}}
\ead{sdas@mst.edu}
\fntext[label1]{the University of Texas at Arlington}
\fntext[label2]{Missori University of Science and Technology}

\begin{abstract}
Wireless Sensor Networks (WSNs) are being deployed for different applications, each having its own structure, goals and requirements. Medium access control (MAC) protocols play a significant role in WSNs and hence should be tuned to the applications. However, there is no for selecting MAC protocols for different situations. Therefore, it is hard to decide which MAC protocol is good for a given situation. Having a precise model for each MAC protocol, on the other hand, is almost impossible. Using the intuition that the protocols in the same behavioral category perform similarly, our goal in this paper is to introduce a general model that selects the protocol(s) that satisfy the given requirements from the category that performs better for a given context. We define the Combined Performance Function (CPF) to demonstrate the performance of different categories protocols for different contexts. Having the general model, we then discuss the model scalability for adding new protocols, categories, requirements, and performance criteria. Considering energy consumption and delay as the initial performance criteria of the model, we focus on deriving mathematical models for them. The results extracted from CPF are the same as the well-known rule of thumb for the MAC protocols that verifies our model. We validate our models with the help of simulation study. We also implemented the current CPF model in a web page to make the model online and useful.  
\end{abstract}


\end{frontmatter}

\section{Introduction} \label{sec:intro}
Unique characteristics of wireless sensor networks (WSNs), in addition to being mostly application-specific, make traditional network algorithms and protocols unsuitable for them. Specifically, some characteristics of WSNs are as follows: (i) wireless sensor nodes usually have limited resources such as available energy, storage, computation and communication capabilities; (ii) the amount of data transmitted is typically lower than other networks (e.g., Wi-Fi); and (iii) wireless links are unreliable by nature, with an additional caveat that nodes spend considerable time in the sleep state to save energy.

We also note that the characteristics of sensor networks may be different in different contexts. For example, small sensor networks used in farming have fewer nodes with more resources \cite{li2011wireless}; traffic load may be significantly higher in multimedia sensor networks \cite{poojary2010multipath}; links are more unreliable in underwater sensor networks \cite{chao2013multiple}; whereas at the other extreme, in some WSNs (e.g. the floating sensors project \cite{wu2009data}), cell phones are used as sensor nodes and the cellular network provides a centralized infrastructure for communication.

In most WSNs, the medium access control (MAC) sub-layer provides mechanisms and policies for sharing the wireless medium. Clearly, not all MAC protocols are well suited for every situation. MAC protocols for WSNs can be classified in several ways. Some survey articles~\cite{Networking2005,demirkol2006mac,kredo2007medium} have focused on the traditional taxonomy. However, these classifications do not take the application context of individual sensor networks into account, and hence provide only limited insights. The authors in \cite{bachir2010mac} classify MAC protocols based on their behavior and claim that each category is useful for a different traffic load. For example, in high traffic, scheduled protocols are said to perform better, because they use pre-scheduling to prevent collision and reduce overhearing and idle listening. Similar behavioral categorization is depicted in \cite{huang2013evolution} by showing the evaluation of MAC protocols for wireless sensor networks over the period 2002-2011.

Before designing and deploying a WSN, an important question often arises: which MAC protocol is better for a given situation? Since there is a lack of unified analytical models that analyze the behavior of MAC protocols under different conditions, it is hard to address this question satisfactorily. Thus, most decisions are made based on questionable "rule of thumb" engineering principles. One may say using the known rule of thumb is enough for making the decision. Example~\ref{example:1} explains two scenarios that show how difficult such task may be.

\begin{example} \label{example:1}
Suppose we are looking for a MAC protocol for an environment-monitoring application with the specifications and the network characteristics mentioned in Table \ref{table:notations} (Except for the number of nodes, network radius, and packet generation rate).
For the security reasons, the MAC protocol should prevent overhearing; moreover, being independent from the network topology, we are looking for a protocol with distributed manner.
Based on the application, energy consumption is the main concern; however the delay also should be reasonable.
Consider the following two scenarios. In the first scenario there are 90 nodes distributed uniformly on a field with radius 100 and average network packet generation rate is 100 packets per second. The network in the second scenario contains 110 nodes and the network radius is 70.
\end{example}

We will show in Section~\ref{sec:cpf} that even slight changes may greatly affect the performance of MAC protocols. For each scenario in Example~\ref{example:1}, we will also select a MAC protocol based our current protocol pool and the model we propose in this paper.

On the other hand, the number of proposed MAC protocols for WSNs is large (and still rising); this makes it almost impossible to obtain a precise analysis for each one of them. Intuitively, the protocols in the same behavioral category have similar performances. Therefore, if we can decide which category is better for a given situation, we can use a qualitative comparison to find the best match. Using this intuition, we introduce a general model for selecting MAC protocols for the wireless sensor networks. We try to make the model scalable so that new categories, protocols, requirements, and performance criteria can be added to it gradually.

In a related work \cite{langendoen2010analyzing}, the authors analyze the performance of low data-rate WSNs. While they focus on low data-rate settings, in this work we aim to produce a more general model that applies not only to traditional low data-rate WSNs, but also more recent WSNs featuring higher traffic loads (e.g., multimedia WSNs). As far as low data-rate WSNs are concerned, our results should further validate the results presented in \cite{langendoen2010analyzing}. There seems to be general consensus that, using the rule of thumb is not sufficient for selecting the MAC protocols and adapting their parameters; for example \cite{zimmerling2012ptunes} argues along the same line when presenting pTunes, where the base station selects the best protocol (among X-MAC~\cite{buettner2006x} and Koala~\cite{liang2008koala}) based on network feedback.

Our contributions are summarized as follows:
\begin{itemize}
\item The main contribution of the paper is the introduction of a general model for selecting MAC protocols for wireless sensor networks for different network specifications and protocol settings, requirements, and performance criteria importance/cost functions. Our model helps find the protocol(s) that satisfy the requirements from the category that performs better for the given situation.
\item We define the Combined Performance Function to compound the performance analysis under different criteria.
\item We show how new protocols, categories, and the requirements can be added to the model, making our model future proof.
\item Mainly focusing on performance analysis, we consider energy consumption and end-to-end delay as the initial performance criteria and derive the mathematical performance model for the three categories of MAC protocols mentioned in \cite{bachir2010mac}.
\end{itemize}

Based on a rule of thumb, we expect that preamble sampling protocols are well suited for low traffic environments, common active period protocols offer better performance in medium traffic situations, while scheduled protocols act well at low node population and high traffic loads. We will show in Section V that these rules of thumb strongly correlate with the findings based on our model.  We also validate our models by performing detailed simulation studies. The initial version of our model with a web user interface is accessible online.

The rest of the paper is organized as follows. The general model is presented in Section II, including the Combined Performance Function (CPF) and the description of model scalability. Section III develops energy consumption models used in our analysis. Approximate delay models are derived in Section IV. Simulation results are presented in Section V to validate our models. Finally, conclusions are offered in Section VI.

\section{General Model}
In this section we present our general model for MAC protocol selection. The intuition behind our model is that if MAC protocols are behaviorally clustered, then protocols in the same category should have similar performance characteristics. Using the categorization presented in \cite{bachir2010mac}, Table~\ref{table:1} shows a qualitative comparison between MAC protocols of different categories listing major behavioral characteristics that affect their performance. Although Table~\ref{table:1} does not provide quantitative performance data, it indicates that protocols in the same category have similar characteristics. We use this observation to create a MAC protocol selection framework that simply removes all the categories and do not have any protocols satisfying the requirements,then uses a performance metric for ranking the remaining categories, and selects a satisfying protocol from the best category.

\begin{table}
\begin{scriptsize}
\begin{center}
\begin{tiny}
\begin{tabular}{|@{}c@{}|@{}c@{}|@{}c@{}|@{}c@{}|@{}c@{}|@{}p{0.9cm}@{}|@{}p{0.7cm}@{}|@{}p{0.7cm}@{}|@{}l@{}|}
 \hline
 \textbf{Protocol}&\textbf{Category}&\textbf{Manner}&\textbf{Scalable}&\textbf{Delay}&\textbf{Collision-free}&\textbf{Idle listening}&\textbf{Over hearing}&\textbf{Overhead} \\ \hline
TSMP \cite{tsmp}&\tobj{ScP}&Cent.&No&Long&No&short&No&Synchronization, Control messages, duty cycling, Timing error \\ \hline
Arisha \cite{arisha}&\tobj{ScP}&Cent.&No&Long&No&short&No&Synchronization, Control messages, duty cycling, Timing error \\ \hline
GinMAC \cite{ginmac}&\tobj{ScP}&Cent.&Yes&Long&No&short&No&Synchronization, Control messages, duty cycling, Timing error \\ \hline
SMACs \cite{smacs}&\tobj{ScP}&Dist.&Yes&Long&Yes&short&No&Scheduling, Synchronization, Control messages, duty cycling, Timing error \\ \hline
Pedamacs \cite{pedamacs}&\tobj{ScP}&Cent.&No&Long&Yes&short&Yes&Setup Phase, Synchronization,Control messages, duty cycling, Timing error \\ \hline
AS-MAC \cite{asmac}&\tobj{ScP}&Dist.&Yes&Long&Yes&Yes&Yes&Setup Phase, Synchronization, Control messages, duty cycling, Timing error \\ \hline
SMAC \cite{smac}&\tobj{CAP}&Dist.&Yes&Medium&Yes&Yes&Yes&Control Messages, Synchronization, duty cycling \\ \hline
TMAC \cite{tmac}&\tobj{CAP}&Dist.&Yes&Medium&Yes&Yes&Yes&Control Messages, Synchronization, duty cycling \\ \hline
NanoMAC \cite{nanomac}&\tobj{CAP}&Dist.&Yes&Short&Yes&Yes&Yes&Control Messages, Synchronization, duty cycling \\ \hline
UMAC \cite{umac}&\tobj{CAP}&Dist.&Yes&Medium&Yes&Yes&Yes&Control Messages, Synchronization, duty cycling \\ \hline
MSMAC \cite{msmac}&\tobj{CAP}&Dist.&Yes&Medium&Yes&Yes&Yes&Control Messages, Synchronization, duty cycling \\ \hline
QMAC \cite{qmac}&\tobj{CAP}&Dist.&Yes&Medium&Yes&Yes&Yes&Control Messages, Synchronization, duty cycling \\ \hline
CL-MAC \cite{clmac}&\tobj{CAP}&Dist.&Yes&Medium&Yes&Yes&Yes&Control Messages, Synchronization, duty cycling \\ \hline
PSA \cite{psa}&\tobj{PSP}&Dist.&Yes&Short&Yes&short&short&Preamble Overhead, duty cycling \\ \hline
BMAC \cite{bmac}&\tobj{PSP}&Dist.&Yes&Short&Yes&short&short&Preamble Overhead, duty cycling \\ \hline
STEM \cite{stem}&\tobj{PSP}&Dist.&Yes&Short&Yes&short&No&Control Messages, Preamble Overhead, duty cycling \\ \hline
MH-MAC \cite{mhmac}&\tobj{PSP}&Dist.&Yes&Short&Yes&short&short&Preamble Overhead, duty cycling \\ \hline
DSP-MAC \cite{dspmac}&\tobj{PSP}&Dist.&Yes&Short&Yes&short&short&Beacon and Control Messages, duty cycling \\ \hline
RICER \cite{ricer}&\tobj{PSP}&Dist.&Yes&Short&Yes&short&short&Preamble Overhead, duty cycling \\ \hline
WiseMAC \cite{wisemac}&\tobj{PSP}&Dist.&Yes&Short&Yes&short&short&Synchronization, Preamble Overhead, duty cycling \\ \hline
RI-MAC \cite{rimac}&\tobj{PSP}&Dist.&Yes&Short&Yes&long for sender&Yes for sender&Beacon Overhead, duty cycling \\ \hline
X-MAC \cite{buettner2006x}&\tobj{PSP}&Dist.&Yes&Short&Yes&short&short&Preamble Overhead, duty cycling \\ \hline
Koala \cite{liang2008koala}&\tobj{PSP}&Dist.&Yes&Short&Yes&long for sender&Yes&Preamble Overhead, Probe and Ack., duty cycling \\ \hline
CLOA\cite{cloa}&\tobj{PSP}&Dist.&Yes&Short&Yes&short&Yes&Bacon Overhead, duty cycling \\ \hline
A-MAC \cite{amac}&\tobj{PSP}&Dist.&Yes&Short&Yes&short&short&Probe overhead, auto Ack. frame, P-CW, duty cycling \\ \hline
\end{tabular}
\end{tiny}
\end{center}
\end{scriptsize}
\vspace{-3mm}
\caption{Qualitative comparison of important existing MAC protocols for wireless sensor networks.} 
\label{table:1}
\end{table}

Algorithm~\ref{alg:basic} presents the simple MAC protocol selection framework for a given context, where $\xi$ is network specifications and protocols settings, $R$ is application requirements, and $\varsigma$ is importance/cost function. It determines the categories that have at least one protocol which satisfies the requirements $R$. Note that a protocol-table that shows which protocols satisfy which set of requirements (e.g. mobility, robustness, scalability, and security) is required (for example in Example~\ref{example:1}, the requirements are security(over hearing prevention) and having a distributed manner). The algorithm then computes the performance of each category using the CPF (see Section~\ref{subsec:cpf}) and finds the category $C_{opt}$ that has maximum performance for the context and provided coefficients. Finally it returns the protocols in the optimal category that satisfy the requirements.

\begin{algorithm}[!ht]
\small
\LinesNumbered
\SetCommentSty{textsf}

\KwIn{
\begin{itemize}
\item $\xi$: network specifications and protocols settings
\item  $R$: application requirements
\item $\varsigma$: importance/cost function
\end{itemize}
}

\KwOut{best matching protocol $p_{opt}$}

\BlankLine
$\Psi \leftarrow \{category\; C | \exists p\in C\, s.t.\, \forall r\in R:\,r[p]=true\}$\;

\ForEach{$C\in \Psi$}
{
	$C.\eta \leftarrow CPF(C,\xi,\varsigma)$\;
}
$C.\eta \leftarrow CPF(C,\xi,\varsigma)$\;
$C_{opt}\leftarrow findMaxPerformance(C)$\;
$p_{opt}\leftarrow \{p\in C_{opt}\, s.t.\, \forall r\in R,r[p]=true\}$\;

\BlankLine
\Return $p_{opt}$\;
\caption{MAC protocol selection framework}
\label{alg:basic}
\end{algorithm}

\subsection{Combined Performance Function} \label{subsec:cpf}
The performance of a category of behaviorally similar protocols is defined as the combination of the mathematically analyzed performance of the representative protocol (the (pioneer) protocol that shows the general behavior of the category) of each category under different criteria. Therefore, we define a Combined Performance Function ($CPF$) that combines the models under different criteria into a single scalar measure based on which the best category of MAC protocols for a given context is selected. The criteria that have direct effect on the performance are placed in the numerator $N$ and criteria that have an inverse effect on the performance will be placed in the denominator $D$. Some may say combining different criteria by summation is like adding apple and orange. Thus, in order to combine the values of different measures, we need a cost function $\kappa$  for each measurement. Moreover, different criteria may have different importance in each application. For example, delay may be more important than energy consumption for a fire detection sensor network. Hence the combined performance function has to take the importance $\rho$ of each criterion for the given application into account. We can now define the $CPF$ as follows:
$$ CPT = \frac{\underset{\forall N_i \in N}{\sum} \rho_{N_i} \times \kappa_{N_i} \times N_i}{\underset{\forall D_i\in D}{\sum} \rho_{D_i} \times \kappa_{D_i} \times D_i} $$
Due to the nature and the application scenarios of wireless sensor networks, energy consumption and delay are two of the most important criteria. Thus we selected them as the current performance criteria for the model and we will show detailed analysis over them in the sections 3 and 4. Note that other criteria can also be added to the model later as is explained in subsection 2.2 and the presented model is agnostic to the selected criteria. We assume that κ and ρ are linear functions and denote $\alpha = \rho_E \times \kappa_E$ and $\beta = \rho_{T_\delta} \times \kappa_{T_\delta}$

as the importance/cost coefficient for energy consumption and delay respectively; where $E$ represents the energy consumption model and $T_\delta$ is the delay model. Both energy consumption and delay have inverse effect to the performance. The $CPF$ therefore for delay and energy consumption becomes:
$$CPF=\frac{1}{\alpha E + \beta T_\delta}$$
\subsection{Model Expansion}
As mentioned earlier, the number of research endeavors in sensor networks is enormous and new MAC protocols are introduced with high frequency. In this paper, we consider the behavioral categorization presented in \cite{bachir2010mac}. 
Due to the large research interest in sensor networks, we cannot possibly mention and include all protocols, requirements, or criteria that are not covered in this paper or some may be discovered later. There also is a chance that there are (will be) the protocols that does not belong to the current categories. 

\begin{figure}
    \centering
    \includegraphics[width=105mm]{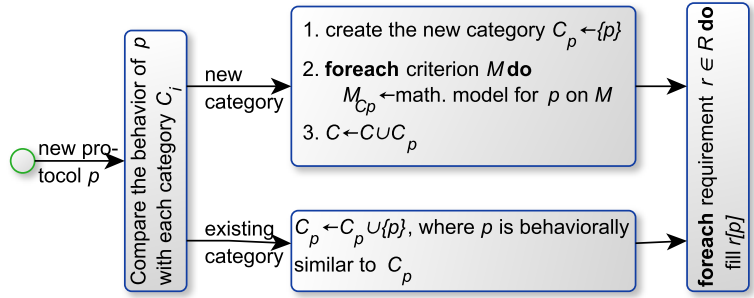}
    \caption{\small{Model expansion with a new protocol.}}
    \label{fig:NewProtocol}
    \vspace{-2mm}
\end{figure}

One of the more important features that the model has to have is expandability so that new protocols, categories, requirements, and performance criteria can be added progressively. In this section we focus on this aspect and explain how the model can be expanded; Figure~\ref{fig:NewProtocol} shows an outline on adding a new protocol or a new category to the model.

Adding a new category to the model requires the analysis of its representative protocol for every performance criterion in the model. However, considering that the current behavioral categorization of MAC protocols is relatively comprehensive, there is likely only a few categories that will surface and need to be added to the model. Adding a new performance criterion requires precise analysis of the representative protocols of each category.

Adding a new requirement to the model compels a review of all included MAC protocols to check whether they satisfy the requirement; this means that all protocols in the model repository need to be checked. We acknowledge that this can be a daunting task but we argue that new requirements surface with a significantly lower frequency than new MAC protocols. However, the following heuristic can be applied for such cases. Given that we are interested in the protocols that satisfy the application requirements, we can classify the protocols of each category based on the combination requirements they satisfy; then we can select a set of protocols that cover the maximum combinations and check if they satisfy the new requirements and continue to update the set until we get a set of requirements that satisfy the new requirement and their collection covers the maximum combination of current requirements.

In sections 3 and 4, we will analyze the representative protocols of the scheduled protocols (Time Synchronized Mesh Protocol (TSMP)), common active period protocols (Sensor MAC (SMAC)), and preamble sampling protocols (Preamble Sampling Aloha (PSA)) for the two current performance criteria, energy consumption and delay, of the model respectively.
Some may ask why did we select these protocols (rather than the more recent/advanced protocols). We had two main reasons: (i) since most of the more recent/advanced protocols are improvement on the basic protocol of their category, the basic protocol may present their common features better, (ii) rather than complicating the analysis, we wanted to make them simpler.
Table~\ref{table:notations} summarizes the notations and the default values used in the analysis.

\begin{table}
\begin{scriptsize}
\begin{center}
\begin{scriptsize}
\begin{tabular}{|@{}c@{}|@{}c@{}|@{}c@{}|@{}c@{}|}
 \hline
 &\textbf{Notation}&\textbf{Meaning}&\textbf{Default value} \\ \hline
\multirow{4}{*}{General} &$L_m$&Message length&4000 bits \\ \cline{2-4}
&$L_h$&Control messages length&240 bits \\ \cline{2-4}
&$d$&Transmission range&20 m \\  \cline{2-4}
&$R$&Network radius&100 m \\  \cline{2-4}
&$N$&Number of nodes&100 \\  \cline{2-4}
&$G$&Network packet generation rate&20 $\frac{1}{sec}$ \\  \cline{2-4}
&$B$&Bandwidth&256000 $\frac{bit}{sec}$ \\  \cline{2-4}
&$\Delta$&Node density&0.003183099 $\frac{1}{m^2}$ \\ \hline
\multirow{5}{*}{Energy}&$P_{Idle}$&Power consumption in idle state&0.003 W \\  \cline{2-4}
&$E_{on}$&Required energy to activate the node&0.000003 J \\  \cline{2-4}
&$E_{off}$&Required energy to deactivate the node&0.000003 J \\  \cline{2-4}
&$E_{send}(d)$&Required energy for transmitting 1 bit with range $d$&0.0000003 J \\ \cline{2-4}
&$E_{rcv}$&Required energy for receiving 1 bit&0.00000003 J \\ \hline
\multirow{3}{*}{TSMP} &$T_g$&Timing error tolerance&0.002 sec \\  \cline{2-4}
&$T_{slot}$&Length of a slot&0.02753125 sec \\  \cline{2-4}
&$T_f$&Length of a super frame&3.670833333 sec \\ \hline
\multirow{4}{*}{SMAC} &$dc$&Duty cycling active period&0.3 sec \\  \cline{2-4}
&$CW_{min}$&Minimum size of collision window&0.00001 sec \\  \cline{2-4}
&$CW_{max}$&Maximum size of collision window&0.001 sec \\  \cline{2-4}
&$M$&Number of increase to $CW_{max}$&6 \\ \hline
\multirow{3}{*}{PSA} &$T_{Interval}$&Channel check period&0.01 sec \\  \cline{2-4}
&$L_p$&Preamble length&4096 bits \\  \cline{2-4}
&$T_{check}$&Channel checking duration&0.000585938 sec \\ \hline

\end{tabular}
\end{scriptsize}
\end{center}
\end{scriptsize}
\vspace{-3mm}
\caption{Notations explanation and the (default) values used for generating the graphs.} 
\label{table:notations}
\end{table}

\section{Energy Model}
Given the bulk of the research and applications about wireless sensor networks, there are many important performance criteria that should be considered for computing the CPF. However, in order to create the initial model, we selected energy consumption and delay, as two of most important performance criteria. We note that other important performance criteria, such as throughput, should be added to the model and the model is agnostic to the performance criteria selection.

Sensor nodes consume energy while acquiring, processing, transmitting, and receiving data. Although energy consumption due to computations is not negligible (e.g., when employing data fusion \cite{luo2011data}), in general, MAC protocols do not incur much computation overhead. On the other hand, since MAC protocols determine physical transmission policies, the largest share of energy consumption is due to transmission/reception of data. Therefore, in the model we focus on the amount of energy consumed for data transmission.
The main factors leading to transmission-related energy consumption include:
\begin{itemize}
\item Collision: nodes use a shared wireless medium that is unreliable, asymmetric with spatio-temporal characteristics. A receiver within the interference range of a transmitting node, while trying to receive from another sender will experience a collision: as a result, the sender and all active nodes in its transmission range, waste energy for transmission and reception of a garbled-up message, respectively.
\item Overhearing: When a sender sends a message to a receiver, all active nodes within its transmission range overhear (receive and decode) the message.
\item Idle Listening: This is resulted from nodes in active reception states while there are no transmissions on the channel.
\item Overhead (Protocol Overhead): the actual payload is not the only component of a transmission instance. MAC protocols introduce additional fields in their protocol header or may even introduce additional control packets which generally is referred as protocol overhead.
\end{itemize}
Relying on the categorization in \cite{bachir2010mac}, the representative protocols of the three current categories (TSMP: scheduled protocols, SMAC: common active period protocols, and PSA: preamble sampling protocols) will be analyzed in this section.

We use s to denote the scheduled protocols, c for the common active period protocols, and p for preamble sampling protocols in the notations. We also use the indices 1, 2, 3, 4 to denote the energy consumption due to collision, overhearing, idle listening, and overhead, respectively. Each category’s energy consumption model would therefore be the summation these four energy usage components:
$$E_k=\underset{i=1}{\overset{4}{\sum}}E_{k_i},\,k\in \{s,c,p\}$$
To have a general framework and to be independent from any specific energy consumption/battery model, in the analysis we use the general terms $E_{send}(d)$ for the amount of energy is required for transmitting 1 bit within range $d$ and $E_{rcv}$ for the required energy required for receiving 1 bit. However, we use the energy model proposed in \cite{goldsmith2005wireless} for producing results and for the experiments. 
\subsection{Scheduled Protocols (\algname{ScP})}
We derive the energy consumption model for a representative scheduled protocol TSMP \cite{tsmp}. TSMP is a centralized protocol that uses prescheduled super frames assigned to pairs of nodes. Each super frame is a table of time division slots and frequency division channels (i.e., slot-frequency cells). More precisely, every cell of the table represents a given time slot and a given frequency which is dedicated to one link between a pair of nodes. No node can have an assigned cell on more than one frequency in the same time slot.

\begin{figure}
    \centering
    \includegraphics[width=65mm]{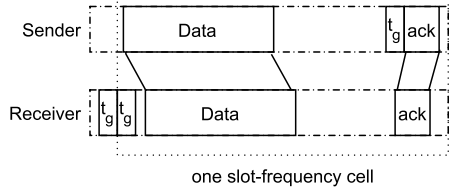}
    \caption{\small{Packet transmission in one slot-frequency cell in TSMP.}}
    \label{fig:TSMP}
    \vspace{-2mm}
\end{figure}

Since each cell is assigned to one link at most, collision is impossible here; and because each node knows its exact wake up and sleep time, there would be no overhearing. However this protocol still suffers from idle listening (because the receiver does not know if there is a packet on the channel and has to stay active in its scheduled rounds) and overhead (Figure~\ref{fig:TSMP}).

\subsubsection{Idle Listening}
Considering the probability of having a packet to transmit in each cell as $Pr$, the average energy consumed for idle listening in a cell is given by $E_{S3_{Cell}}=P_{Idle}(1-Pr)\times T_{Idle}$; where $P_{Idle}$ is the amount of energy used for idle listening per second and $T_{Idle}$ is the amount of time for which the receiver has to stay active to ensure that there is no packet on the channel.

Assuming that network packet generation rate is $G$ packets per second, $G\times T_f$  packets are generated per super frame, where $T_f$ is the length of the super frame in a second. Thus, $Pr=\min(1, \frac{G\times T_f}{N×\times N\prime})$; where $N$ is the number of nodes in the network and $N\prime$ is the number links (neighbors) of a node. Having the transmission range $d$ and the node density $\Delta$, $N\prime =\Delta \times \Pi d^2$.

Every receiver has to listen for $2T_g$ seconds to ensure there is no packet on the channel in this slot (see Figure~\ref{fig:TSMP}). Thus, the total energy ($E_{S3}$) consumed per second for idle listening in network is derived as following equation:
$$E_{S3}=P_{Idle}\times N \times (\Delta \times \Pi d^2)\times [1-\min (1, \frac{GT_f}{N\times (\Delta \times \Pi d^2)} )]×\times 2T_g \times \frac{1}{T_f}$$

\subsubsection{Overhead}
\begin{figure}
        \centering
        \begin{subfigure}[b]{0.33\textwidth}
                \includegraphics[width=\textwidth]{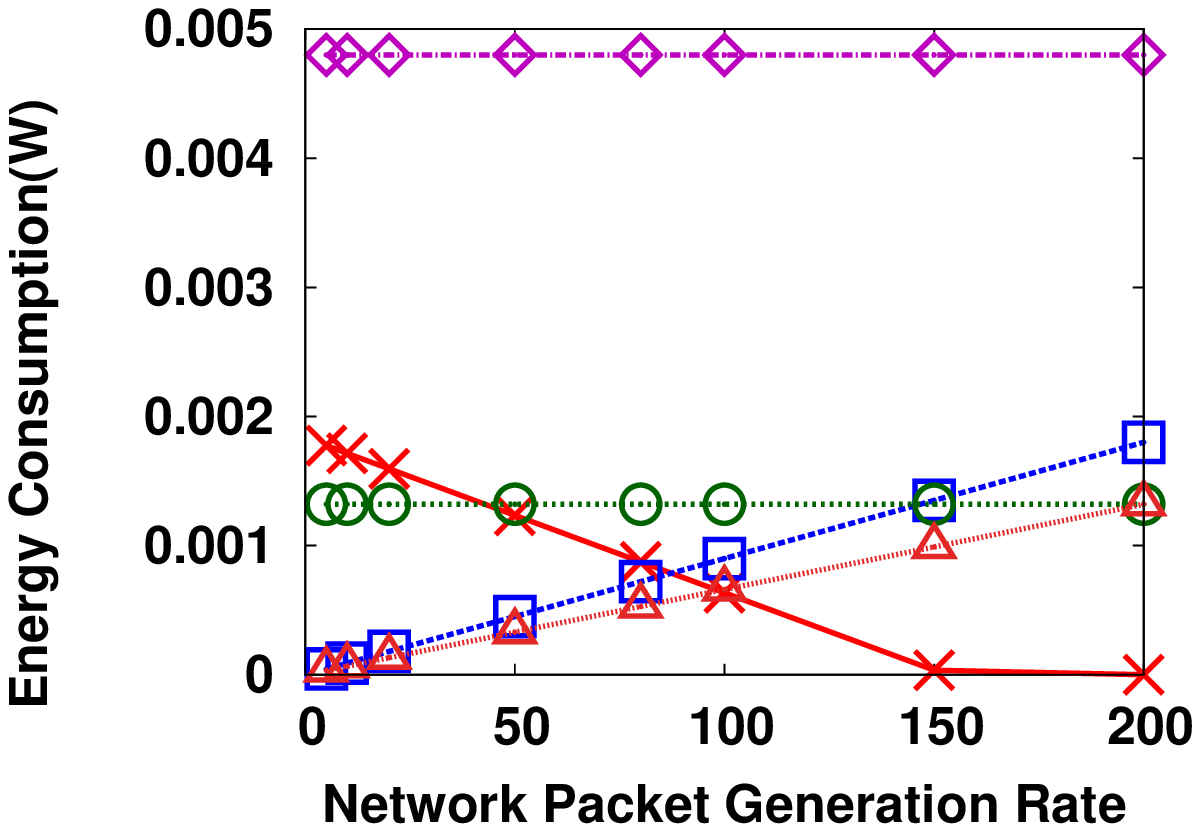}
                \caption{\tiny{$D=20$, $N=100$}}
                \label{fig:TSMPEG}
        \end{subfigure}%
        ~ 
        \begin{subfigure}[b]{0.33\textwidth}
                \includegraphics[width=\textwidth]{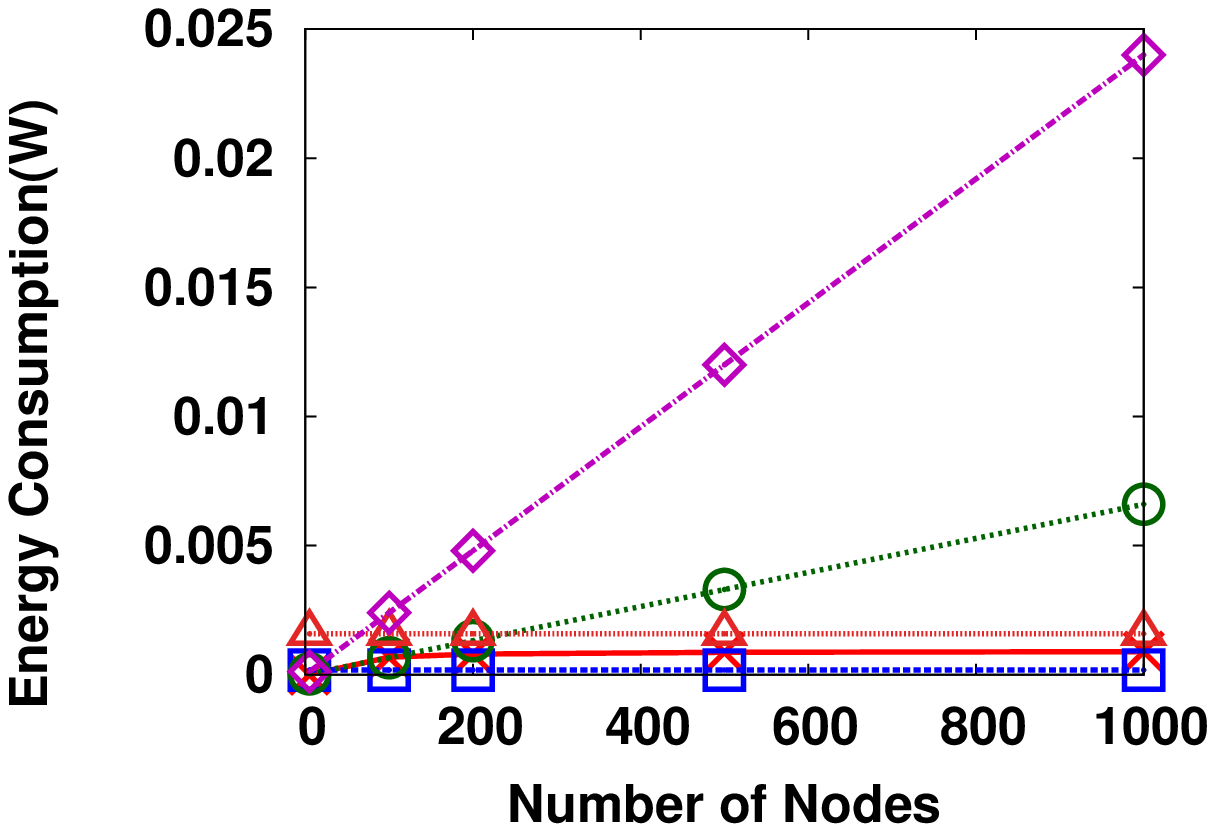}
                \caption{\tiny{$D=20$,$G=20$,$\#ofNeighbors=4$}}
                \label{fig:TSMPEN}
        \end{subfigure}%
        ~
        \begin{subfigure}[b]{0.33\textwidth}
                \includegraphics[width=\textwidth]{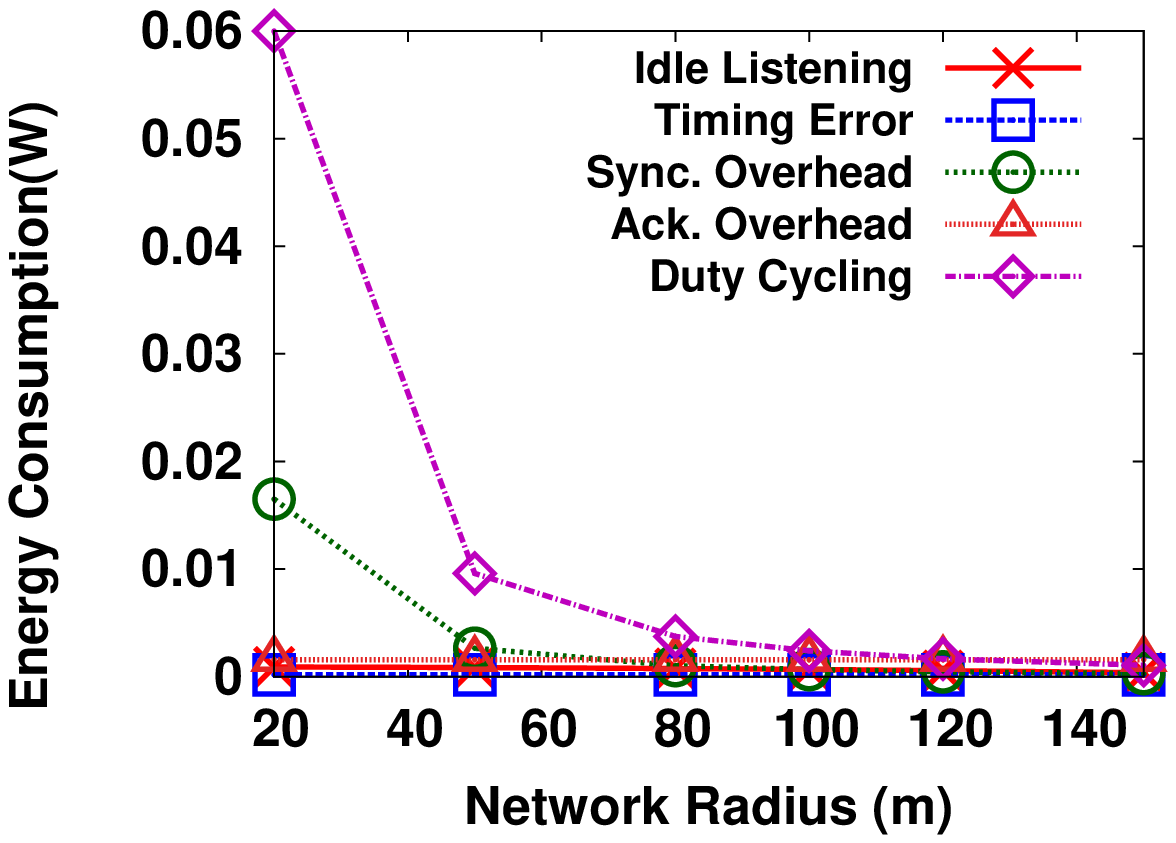}
                \caption{\tiny{$N=100$,$G=20$,$D=20$}}
                \label{fig:TSMPER}
        \end{subfigure}
    \vspace{-5mm}
    \caption{\small{Energy consumption in TSMP for different a) packet generation rates (packet/sec), b) node populations, c) network radii (network density).}}
    \label{fig:TSMP-G}
\end{figure}

Receivers have to wake up $T_g$ seconds before the beginning of their slot. As indicated by Figure~\ref{fig:TSMP} (because nodes may have $T_g$ seconds error in synchronization), the average timing error overhead is $3\frac{T_g}{2}$. Thus the timing error overhead ($E_{S4_1}$) rate can be calculated as
$$E_{S4_1}=P_{Idle}\times G\times \frac{3T_g}{2}$$
To be synchronized with a maximum allowed ($T_g=1$ msec) error, it is enough to send sync packets every 48 seconds \cite{tsmp} and two messages are enough for synchronization \cite{ganeriwal2003timing}. Therefore, the amount of energy ($E_{S4_2}$) used for synchronization overhead is
$$E_{S4_2}=\frac{1}{48}\times 2\times N\times (\Delta\times \Pi d^2)\times (E_{rcv}+E_{send}(d))×L_{Sync}$$
where $L_{Sync}$ is the length of the sync message in bits. Sending and receiving the $ACK$ packets also consume energy ($E_{S4_3}$):
$$E_{S4_3}=G\times L_{Ack}(E_{rcv}+E_{send}(d))$$
where $L_{Ack}$ is the length of the $ACK$ packet. The duty-cycling overhead ($E_{S4_4}$) can be computed 
$$E_{S4_4}=2\times N\times (\Delta \times \Pi d^2)\times (E_{on}+E_{off})$$
Therefore the energy ($E_{S4}$) consumption due to the overhead is derived 
$$E_{S4}=(P_{Idle}\times G\times (\frac{3T_g}{2}+L_{Ack}))+(\frac{1}{48}\times 2\times N\times (\Delta \times \Pi d^2)\times (E_{rcv}+E_{send}(d))$$
$$ \times L_{Sync})+(L_{Ack}(E_{rcv}+E_{send}(d)))+(2\times N\times (\Delta \times \Pi d^2 )×(E_{on}+E_{off} ))$$
Figure~\ref{fig:TSMP-G} is showing the effects of energy consumption under different conditions for TSMP, a category representative for \algname{ScP}. (The values in the graphs are dependent on the properties of sensor nodes and their antenna and are produced based on default values in Table~\ref{table:notations}.) When the number of nodes or the network density increases and consequently the number of links increases, the energy consumption of duty cycling increases. However, since nodes check the channel for a short period of time to ensure it is free, they do not spend a lot of energy for idle listening. Thus, the overall energy consumption is low here.

\subsection{Common Active Period Protocols (\algname{CAP})}
The main idea behind this category of protocols is to reduce the energy consumption due to idle listening (when compared to traditional random access MAC protocols). Nodes have a common schedule according to which they periodically sleep and wake up together. While idle listening and collisions are in trade off, these protocols are not flexible in duty cycling. The representative protocol in this category is SMAC \cite{smac} that uses CSMA/CA random access during active periods. It also uses relative time stamps (rather than absolute) for synchronization; with a recommended sync update message intervals of 10 seconds.

Every newly turned on node listens to packets on the channel to see if there is a schedule being transmitted. If not, it then produces its own schedule and broadcasts it to the network. Nodes with the same sync information form a cluster. Clusters connected by border nodes should work on different schedules to connect virtual clusters together.

Back-off and collision window techniques are used to reduce collision probability and increase network throughput. Since all nodes in a cluster have a common schedule, they all are awake at the same time; and when a node sends a message, all other nodes in the transmission range hear it. Therefore, control messages and duty-cycling are the main overhead resulting in energy loss for this protocol.
\subsubsection{Collision}
We use the collision probability derived in \cite{tickoo2004queueing} for CSMA/CA mechanism; this calculation can be adopted here with some adjustments. Transmissions are initiated with the minimum size collision window of $CW_{min}$; each node waits for a random uniformly distributed back-off time between 1 and $CW$ before a message. Every time a collision occurs (or it is avoided), nodes double the size of $CW$ until it reaches $CW_{max}$. Therefore, as derived in \cite{tickoo2004queueing}, the collision probability is
$$p=1-(1-\frac{\lambda}{\mu}\times \frac{1-2p}{1-p-p(2p)^m} \times \frac{2}{CW_{min}})^{N-1}$$
where $\lambda$ is the packet generation rate, $\mu$ is the service rate in packet per second, $CW_{min}$ is the minimum size of collision window, and m is the number of transmission fails that increases the size of the collision window to $CW_{max}$. 
In this equation $\frac{\lambda}{\mu}$ is the probability that the system is not free, whereas $(\frac{1-2p}{1-p-p(2p)^m} \times \frac{2}{CW_{min}})^{-1}$ is the average window size in a saturated network.

By setting $\lambda=G\times dc$,  $\mu=B$, this formula works here. $dc$ is the duty cycle, i.e., the portion of time that nodes are active together. All the transmissions take place during the active period (which increases the value of $\lambda$). The probability of a successful transmission after $x$ trials is given by $P(x)=(1-p)p^{x-1}$. Thus the expected value of transmissions is $E(x)=\underset{k=1}{\overset{\infty}{\sum}} (1-p)p^{k-1} =\frac{1}{1-p}$ and the average number of collisions for each packet is $\frac{1}{1-p}-1=\frac{p}{1-p}$. 
Therefore, the average energy consumption due to the collision ($E_{C_1}$)  can be obtained 
$$E_{C1}=G\times L_{RTS}\times ((\Delta\Pi d^2-1)E_{rcv}+E_{send}(d))\times \frac{p}{1-p}\times dc$$
where $L_{RTS}$ is the length of the $RTS$ packet. Nodes that overhear the message (a population of $\Delta \Pi d^2$ nodes) and the sender waste energy during the collision. Since a transmission event can only take place during an active period, the above result contains a $dc$ factor.
\subsubsection{Overhearing, Idle Listening, and Overhead}
All nodes in the transmission range of the sender overhear the message. The corresponding energy consumption is              $E_{C2}=L_m\times E_{elec}\times (\Delta \Pi d^2 - 1)\times G$.
Idle listening occurs when the channel is free, however nodes are still listening to it. $G\times \frac{d^2}{R^2}$ is the rate of generated packets  overheard by each node that can be used for determining the average idle listening time in each node. The energy consumption due to idle listening ($E_{C_3}$) is thus:
$$E_{C3}=N\times P_{Idle}\times \max (0,dc-(\frac{L_m+L_{rts}+L_{cts}+L_{ack}}{B}\times \frac{G\times d^2}{R^2}))$$
$\frac{L_m+L_{rts}+L_{cts}+L_{ack}}{B}$ is the amount of time required for transmitting a message (we supposed $L_{rts}=L_{cts}=L_{ack}=L_h$ for producing the graphs and for the experiments).

One \emph{RTS}, one \emph{CTS}, and one \emph{ACK} packets are sent for every message and all nodes in the transmission range of the sender overhear the message. So the overhead of these messages ($E_{C4_1}$) can be derived as
$E_{C4_1}=G\times (L_{rts}+L_{cts}+L_{ack})\times ((\Delta \Pi d^2 -1)E_{rcv} + E_{send}(d))$.

Considering that the sync messages are sent every 10 seconds by every node, since all other nodes in the transmission range of the receiver hear it, the overhead of synchronization ($E_{C4_2}$) is computed as 
$E_{C4_2}=\frac{1}{10} \times ((\Delta \Pi d^2 -1)E_{rcv} + E_{send}(d)) \times L_{sync}\times N$.

Nodes duty cycle once a second to decrease the delay. Each time the nodes sleep and wake up, they spend some energy in the transitions. So the overhead of duty cycling ($E_{C4_3}$) per second in the network is $E_{C4_3}=N\times (E_{on}+E_{off})$. Therefore, the total energy consumed for the overhead is:\\
$E_{C4}=(G\times (L_{rts}+L_{cts}+L_{ack})\times ((\Delta \Pi d^2 -1)E_{rcv} + E_{send}(d)))+N(\frac{L_{sync}}{10} (E_{rcv}(\Delta \Pi d^2 -1) + E_{send}(d)) + E_{on} + E_{off})$

\begin{figure}
        \centering
        \begin{subfigure}[b]{0.33\textwidth}
                \includegraphics[width=\textwidth]{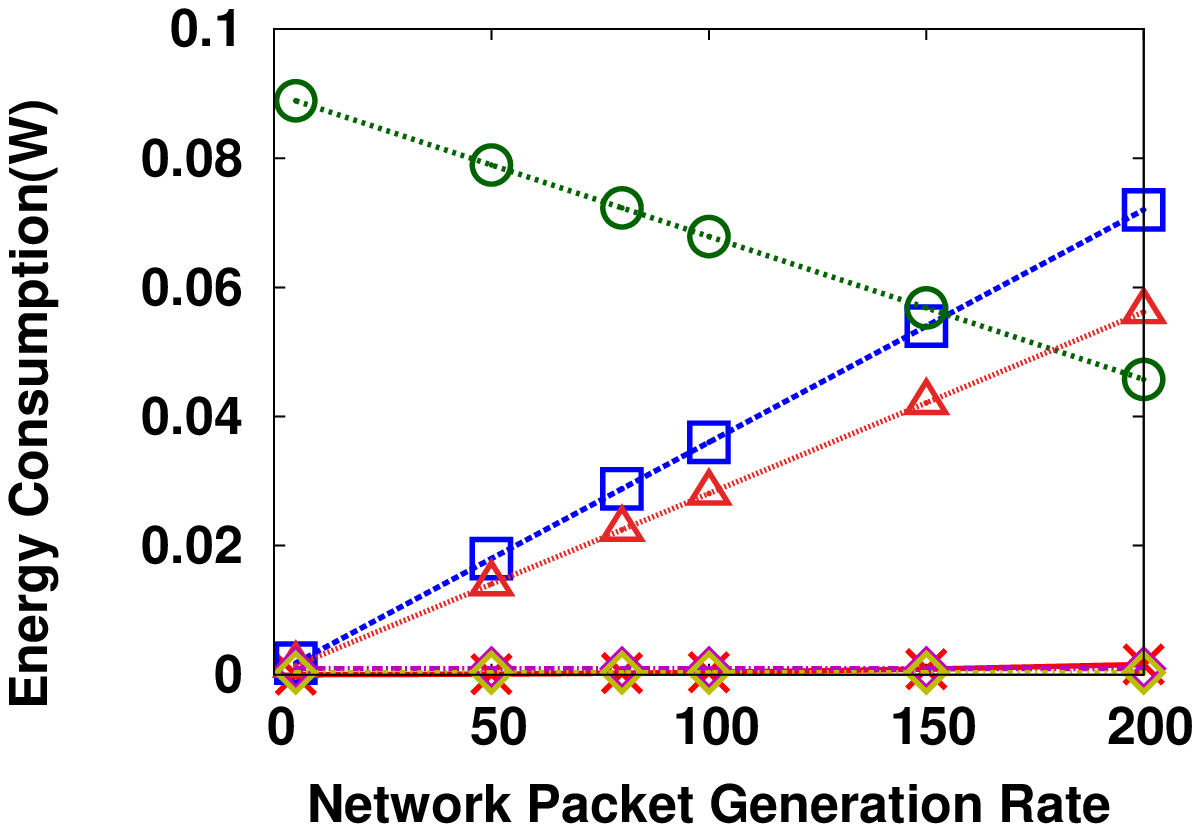}
                \caption{\tiny{$D=20$, $N=100$}}
                \label{fig:SMACEG}
        \end{subfigure}%
        ~ 
        \begin{subfigure}[b]{0.33\textwidth}
                \includegraphics[width=\textwidth]{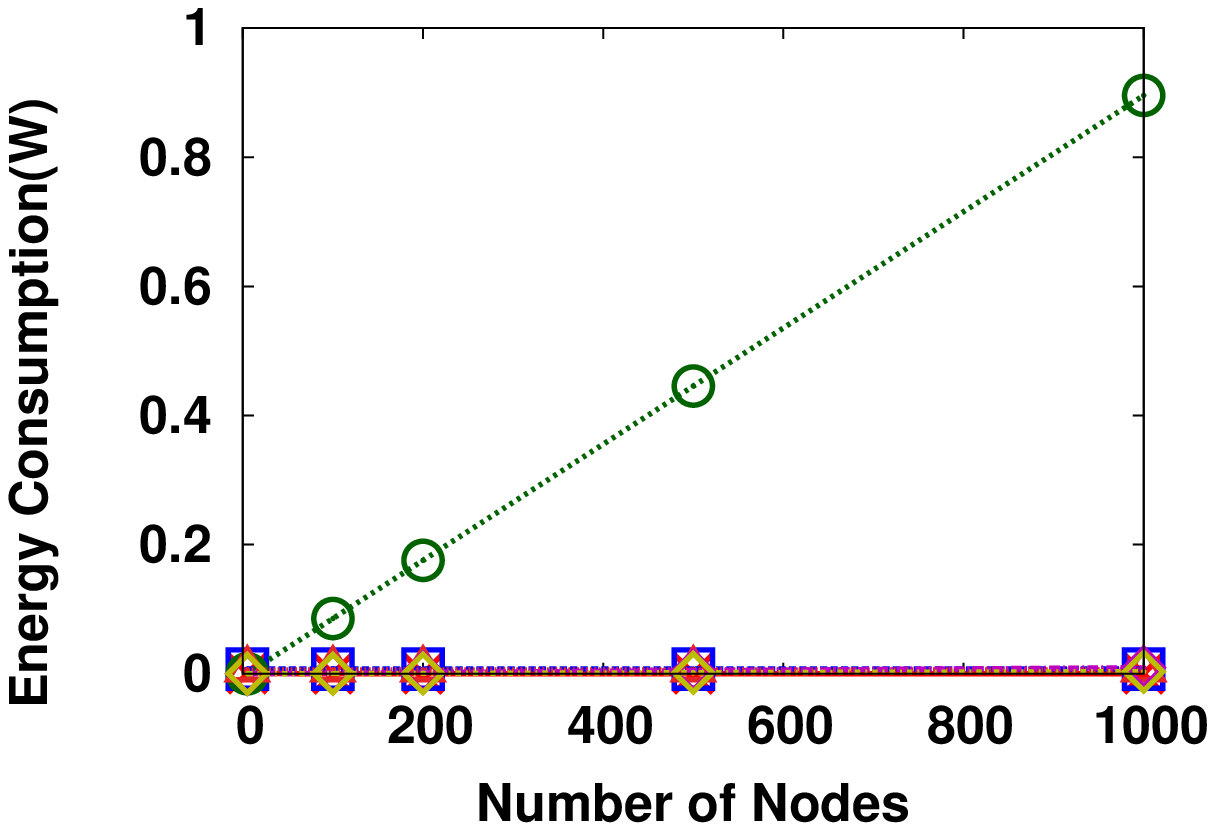}
                \caption{\tiny{$D=20$,$G=20$,$\#ofNeighbors=4$}}
                \label{fig:SMACEN}
        \end{subfigure}%
        ~
        \begin{subfigure}[b]{0.33\textwidth}
                \includegraphics[width=\textwidth]{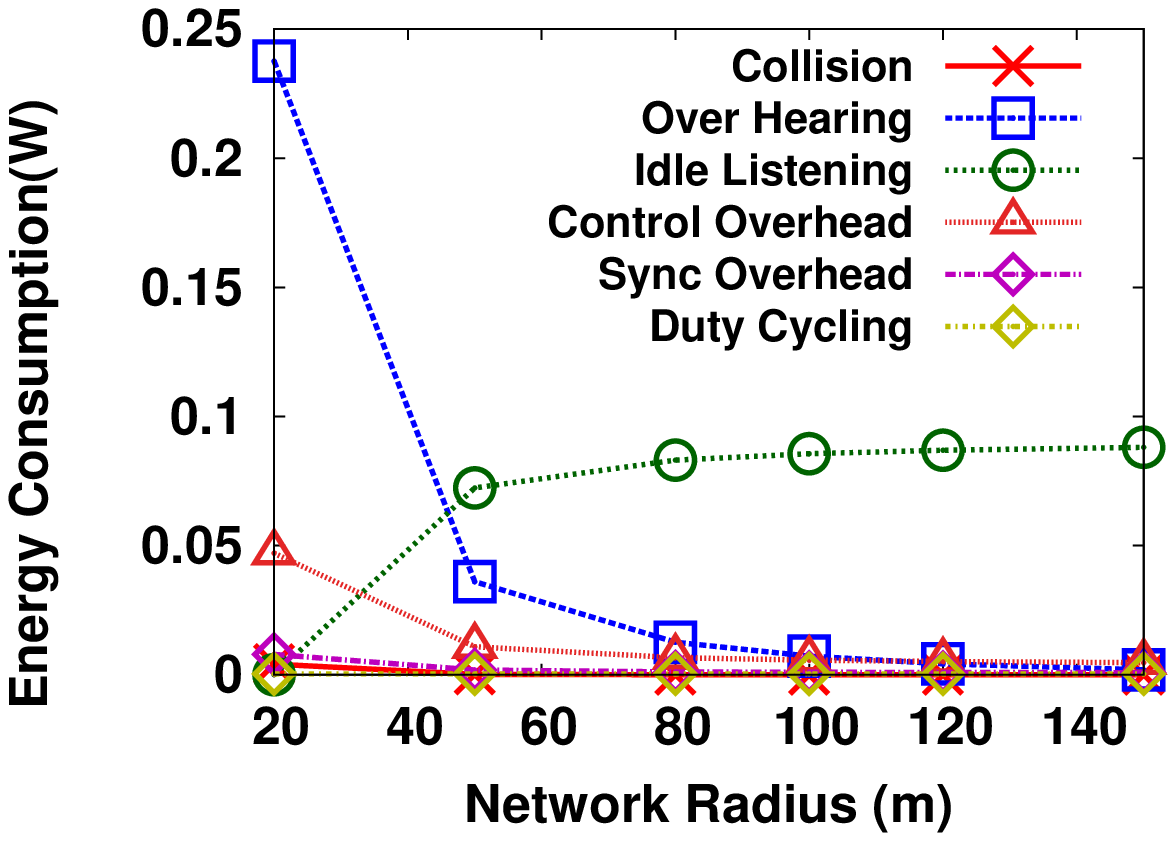}
                \caption{\tiny{$N=100$,$G=20$,$D=20$}}
                \label{fig:SMACER}
        \end{subfigure}
    \vspace{-5mm}
    \caption{\small{Energy consumption in SMAC for different a) packet generation rates (packet/sec), b) node populations, c) network radii (network density).}}
    \label{fig:SMAC-G}
\end{figure}

Figure~\ref{fig:SMAC-G} shows the energy consumption characteristics of SMAC, a representative of common active period protocols. As shown, idle listening and overhearing are the main reasons for energy consumption. This is because the nodes are awake for a long period of time and overhear all the messages that are in their transmission range. 

\subsection{Preamble Sampling Protocols (\algname{PSP})}
\begin{figure}
    \centering
    \includegraphics[width=85mm]{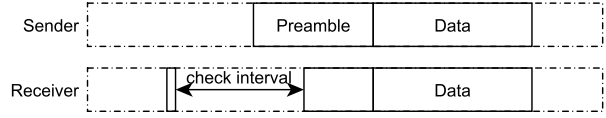}
    \caption{\small{The mechanism of (sender initiated) Preamble Sampling protocols.}}
    \label{fig:PSA}
    \vspace{-2mm}
\end{figure}

In this class of protocols, nodes mostly wake up periodically to check if there is a new message on the channel (Figure~\ref{fig:PSA}). Every node determines its schedule independently. Therefore, synchronization is not required in these protocols. When a node has a message to transmit, first it has to generate a preamble that is long enough to ensure that the intended destination node will receive it at least once ($Preamble\geq Check\_interval$). Since these protocols have a long preamble, collisions are very energy consuming. 

\subsubsection{Collision}
The assumption behind these protocols is that the traffic load (and consequently the collision probability) is low. Here we analyze Preamble Sampling Aloha (PSA) as the representative protocol of this category \cite{psa}. The traffic generation is assumed to follow a Poisson distribution. If the network packet generation rate is G, the packet generation range around each node during the time required for sending the packet is $G\prime=(G\times \frac{d^2}{R^2})\times(\frac{L_p+L_m}{B})$; $\frac{L_p+L_m}{B}$ is the required time for sending a message. No other transmission can be happening in $2\times (transmission\, time)$ in order to have the current transmission successfully completed.
Thus, the probability of generating $x$ messages during a message transmission is $Pr[x]=\frac{e^{-2G\prime}(2G\prime)^x}{x!}$. The probability of a successful transmission after $x$ attempts is given by $P(x)=e^{-2G\prime}\times (1-e^{-2G\prime})^{x-1}$. Therefore, the expected value of transmission attempts is  
$$E(x)=\underset{k=1}{\overset{\infty}{\sum}} k \times e^{-2G\prime}\times (1-e^{-2G\prime})^{k-1} = e^{2G\prime}$$
and the expected value of collision per message is $e^{2G\prime}-1$.

The receiver has to wait for $\frac{L_p}{2}$ seconds on average, before the preamble transmission is finished and data transmission is started. So, the sender has to send $L_p+L_m$ bits for every packet and receiver has to the receive $\frac{Lp}{2}+L_m$ bits. Thus the energy consumption due to collision ($E_{P1}$) in PSA is derived as
$$E_{P1}=(e^{2G\prime}-1)\times (E_{rcv}(\frac{L_p}{2}+L_m)+ E_{send}(d) (L_p+L_m))$$
\subsubsection{Overhearing, Idle Listening, and Overhead}
For a given message, non-destination neighbors overhear $T_{Check}\times B$ bits of preamble during their check interval. Since $T_{Check}$ is small, the energy consumption of overhearing is not significant. The overhearing energy consumption ($E_{P2}$) can be derived as $E_{P2}=T_{Check}\times B \times E_{rcv}\times (\Delta \Pi d^2-1)\times G$.

Idle listening occurs during check intervals, when the channel is unoccupied. The number of channel checks per second is $\frac{1}{T_{Interval}}$. The rate of packets generated in the transmission range of a given node is $\frac{G\times d^2}{R^2}$ . Therefore, every node is in the idle listening mode for $max(0 , (\frac{1}{T_{Interval}}- \frac{G\times d^2}{R^2}))$ seconds. The energy consumption of idle listening ($E_{P3}$) is then 
$$E_{P3}=N\times P_{Idle}\times T_{Check}\times max(0 , (\frac{1}{T_{Interval}}- \frac{G\times d^2}{R^2}))$$
Although $T_{Check}$ is short, since the amount of time that the preamble is in the channel has to be at least equal to $T_{Interval}$, the number of channel checks is high.

\begin{figure}
        \centering
        \begin{subfigure}[b]{0.33\textwidth}
                \includegraphics[width=\textwidth]{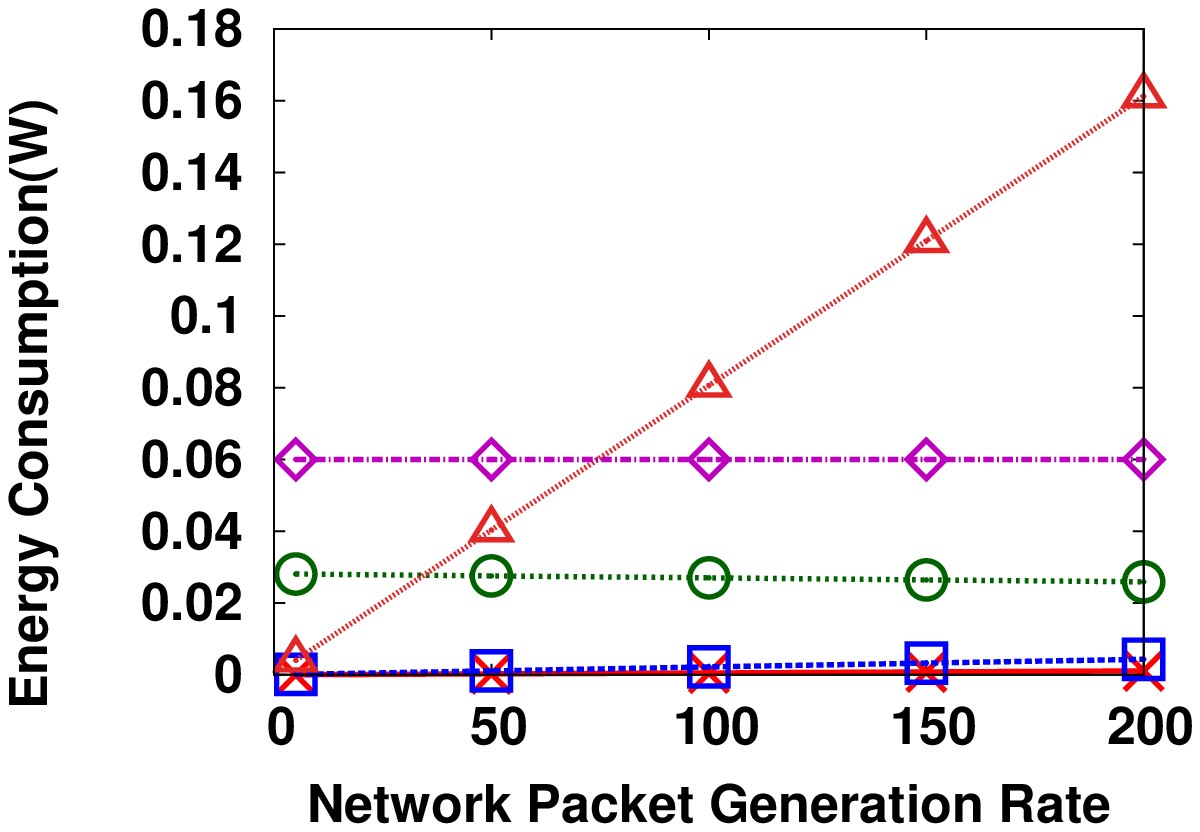}
                \caption{\tiny{$D=20$, $N=100$}}
                \label{fig:PSAPEG}
        \end{subfigure}%
        ~ 
        \begin{subfigure}[b]{0.33\textwidth}
                \includegraphics[width=\textwidth]{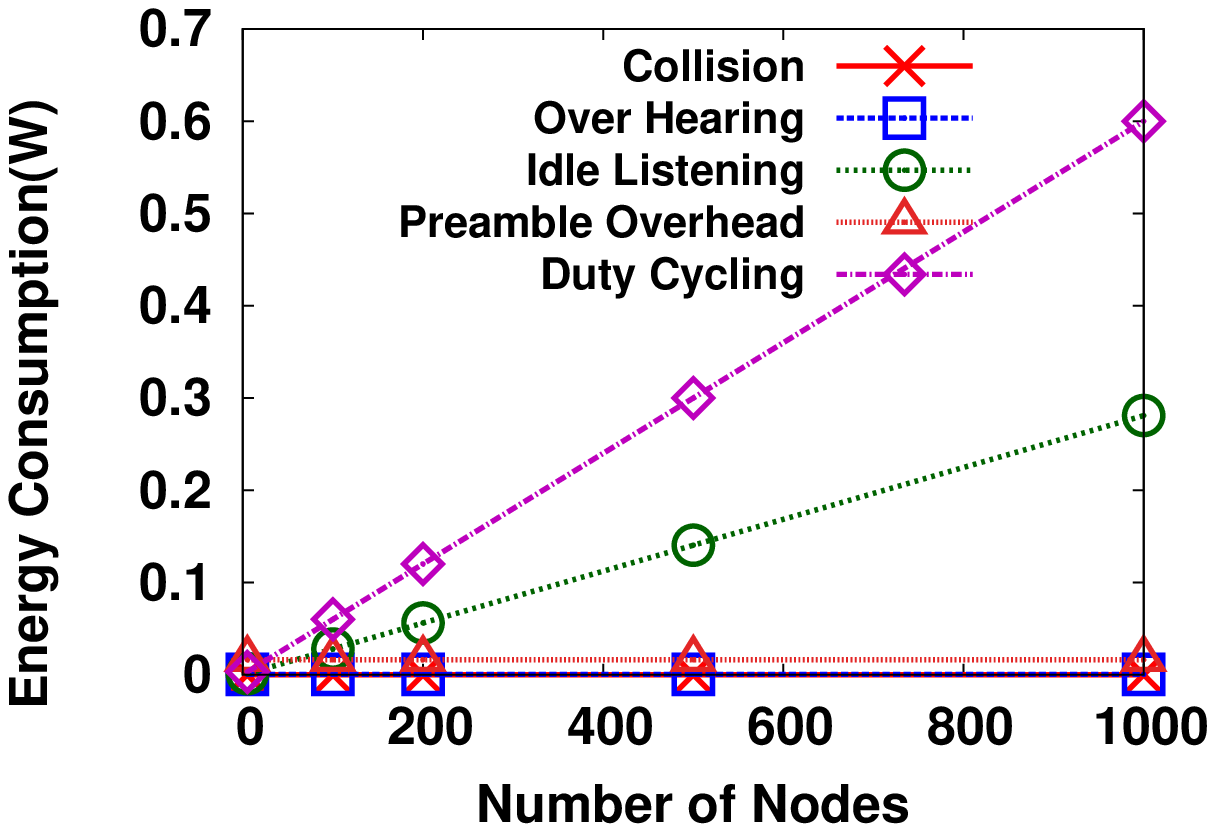}
                \caption{\tiny{$D=20$,$G=20$,$\#ofNeighbors=4$}}
                \label{fig:PSAPEN}
        \end{subfigure}%
        ~
        \begin{subfigure}[b]{0.33\textwidth}
                \includegraphics[width=\textwidth]{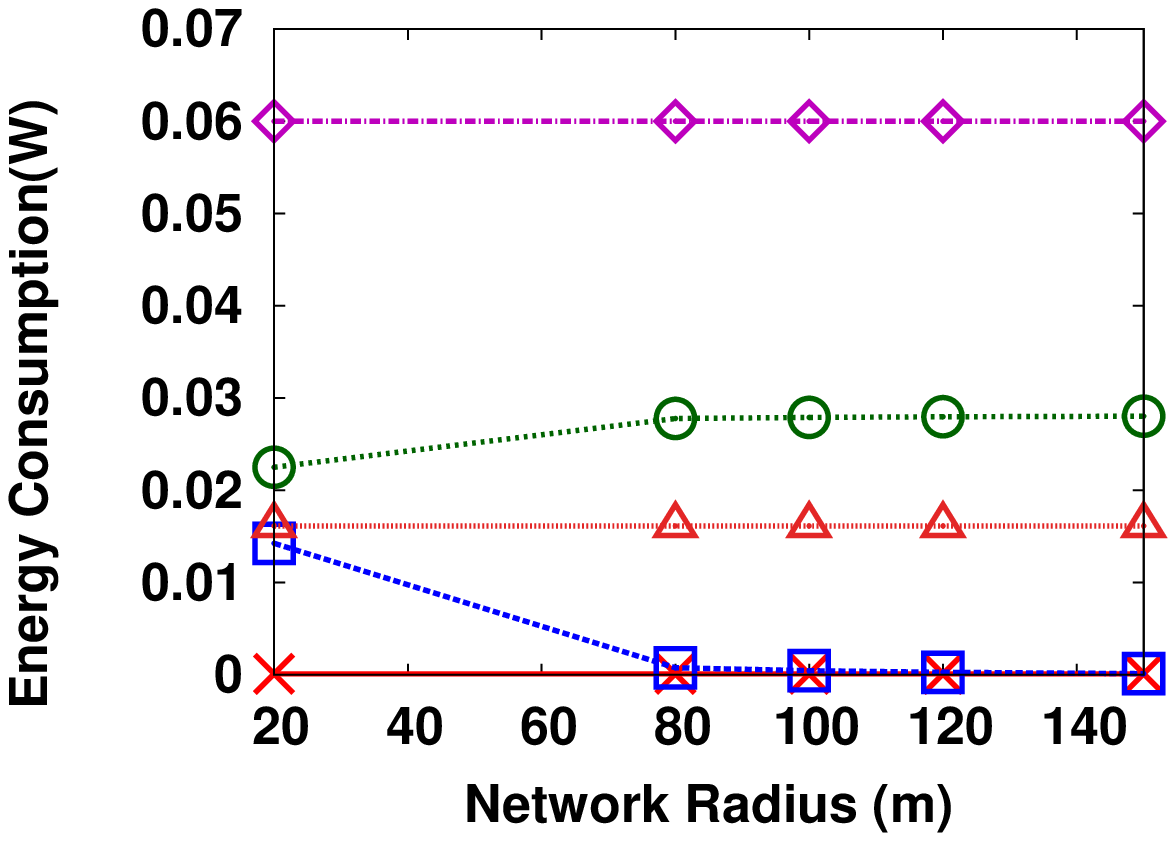}
                \caption{\tiny{$N=100$,$G=20$,$D=20$}}
                \label{fig:PSAPER}
        \end{subfigure}
    \vspace{-5mm}
    \caption{\small{Energy consumption in PSA for different a) packet generation rates (packet/sec), b) node populations, c) network radii (network density).}}
    \label{fig:PSA-G}
\end{figure}

Senders use a long preamble in PSA before sending the message. The receiver also has to listen to  $\frac{L_p}{2}$ bits of preamble, in average. Thus, the overhead of preamble ($E_{P4_1}$) is calculated as $E_{P4_1}= G\times ((\frac{E_{rcv}\times L_p}{2})+ E_{send}(d)\times L_p)$.
The number of check intervals in a second is $\frac{1}{T_{Interval}}$. Therefore, the energy consumption due to the duty cycling ($E_{P4_2}$) overhead is $E_{P4_2}=N\times \frac{1}{T_{Interval}}\times (E_{on}+E_{off})$.  The energy consumption of overhead ($E_{P4}$) is $E_{P4_1}+E_{P4_2}$ such that
$$E_{P4}=\,(G\times ((\frac{E_{rcv}\times L_p}{2})+ E_{send}(d)\times L_p)) \,+ \,(N\times \frac{1}{T_{Interval}}\times (E_{on}+E_{off}))$$
Figure~\ref{fig:PSA-G} shows the energy consumption due to the above reasons under varying network conditions in PSA. Duty cycling, the overhead of preamble transmission, and idle listening are the dominant reasons of energy consumption in this protocol.

\section{Approximate Delay Model}
End to end delay is defined as the time between the instant a packet is passed to the network protocol stack until it gets delivered to the same level protocol (of the protocol stack) in the destination. Modeling such delay is problematic. The problem lies in the “packet generation”; as depending on the complexity of the network model, packets can be generated at the session, network, data link or MAC layers. When investigating MAC protocols, we usually assume a random process that is generating packets to be transmitted by the MAC layer (without modeling the upper layers). Delay in general is the sum of the queuing delays at each layer and the transmission. In our model we only consider the queuing delay at the MAC layer, i.e., from the time a packet is passed to the MAC layer for transmission to the time it is delivered (assuming only a single packet is stored at the MAC layer at any time). In addition we will only look at one-hop delays and thus will not consider the diameter of the network.
\subsection{Scheduled Protocols}
We consider that a packet can be generated any time during the super frame. In the best case the packet is generated exactly at the beginning of its corresponding cell while in the worst case the packet is generated right after the cell belonging to the node has started. Because scheduled protocols are not a random access, the collision probability is zero; i.e., assuming no transmission errors (other than self-interference) occur, it is guaranteed that the packet will be transmitted over the channel successfully in the first upcoming corresponding cell. So the average channel access delay is $\frac{T_f}{2}$ seconds (considering TSMP as the representative protocol of this category). Since the packet is transmitted in one cell, the packet transmission delay is $T_{slot}$. Therefore, the delay can be modeled by
$$T_{\delta _s} = \frac{T_f}{2} + T_{slot}$$
We need to take into account that in the protocols with centralized scheduling, "finding a collision-free schedule is a two-hop coloring problem" \cite{bachir2010mac}. The other issue in scheduled protocols is the size of the super frame. Adding a new node to the network adds several new links (depending on the network density and transmission range), each requiring a specific cell. The size of the super frame is the main reason for delay in this protocol. For example, as mentioned in \cite{tsmp}, "with a 10 ms slot, a cell in a 1000-slot super frame repeats every 10 s".
\subsection{Common Active Period Protocols}
The activity of each node is divided into active and inactive periods in this protocol. The packets generated during the inactive period have to wait until the node is active. On average, the packets generated during the inactive period have to wait for $\frac{1-dc}{2}$ seconds. The portion of packets generated during the inactive period is $1-dc$. As soon as the node becomes active, packets can be transmitted to the destinations.  In this paper we do not consider the number of back-offs into account and assume that the packet is successfully transmitted if the channel is available. Based on Equation (1), the expected number of trials for transmitting a packet is $\frac{1}{1-p}$, where $p$ is the collision probability. The RTS-transmission time is spent for each collision. The packet transmission time has also to be added to the formula. Hence, the approximate delay model for the common active period protocols is given by
$$T_{\delta _c}=(1-dc)\times \frac{1-dc}{2}\, +\, \frac{\frac{L_{RTS}}{1-p}+L_m}{B}$$
\subsection{Preamble Sampling Protocols}
These protocols do not feature carrier sensing and the packet is placed in the channel as soon as it is generated. In addition, even if collision occurs, the sender finishes transmitting the entire packet. We assume there is a feedback informing the sender whether the data have been received. With such feedback, the expected value of trials is calculated with the help of equation (2). The approximate delay model can be derived from the following equation.
$$T_{\delta_p}=e^{2G\prime }\times \frac{L_p+L_m}{B}$$
\section{Combined Performance Function} \label{sec:cpf}
\begin{figure}
        \centering
        \begin{subfigure}[b]{0.33\textwidth}
                \includegraphics[width=\textwidth]{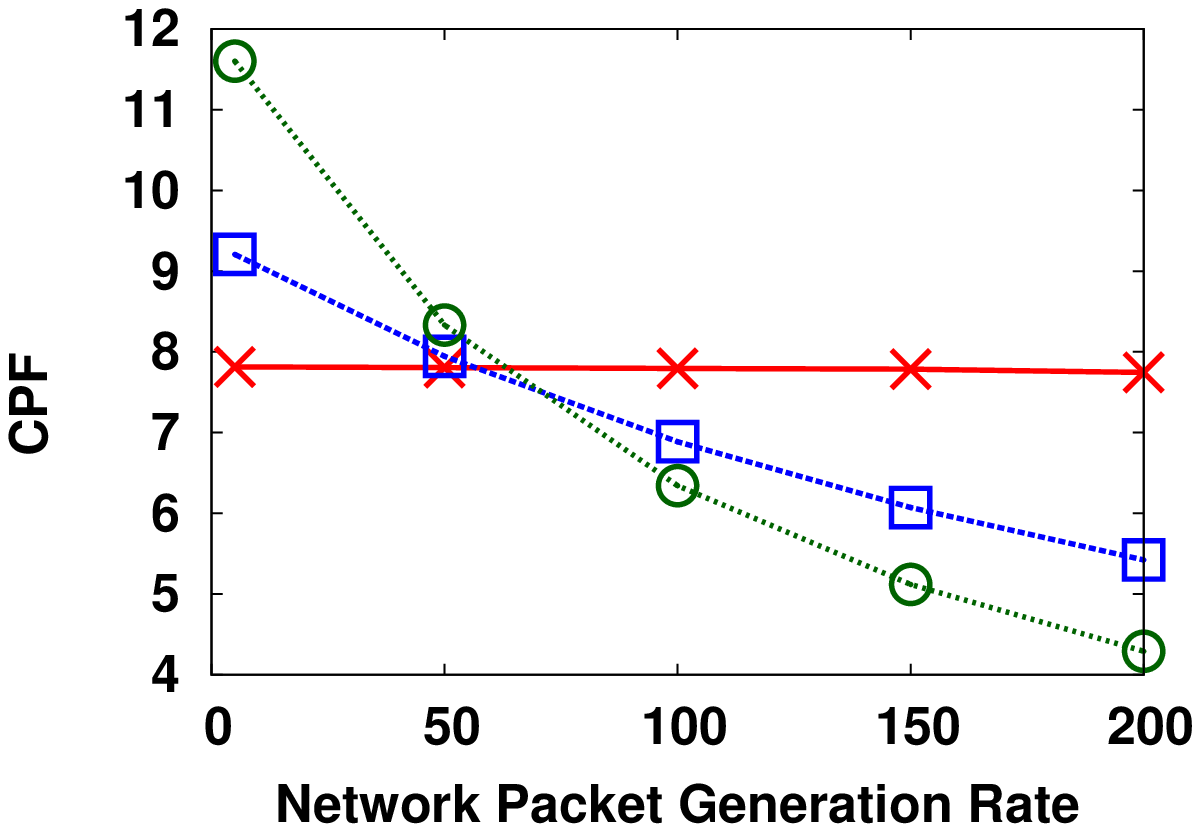}
                \caption{\tiny{ }}
                \label{fig:TSMPEG}
        \end{subfigure}%
        ~ 
        \begin{subfigure}[b]{0.33\textwidth}
                \includegraphics[width=\textwidth]{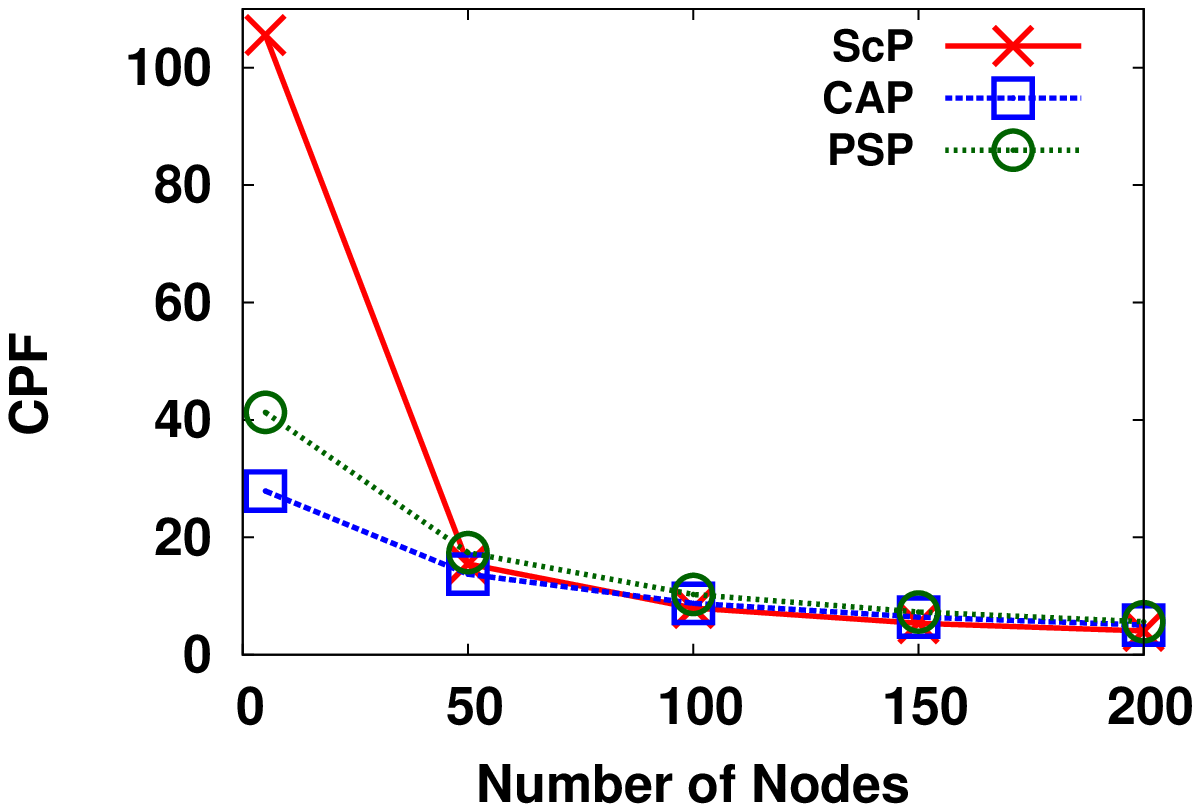}
                \caption{\tiny{ }}
                \label{fig:TSMPEN}
        \end{subfigure}%
        ~
        \begin{subfigure}[b]{0.33\textwidth}
                \includegraphics[width=\textwidth]{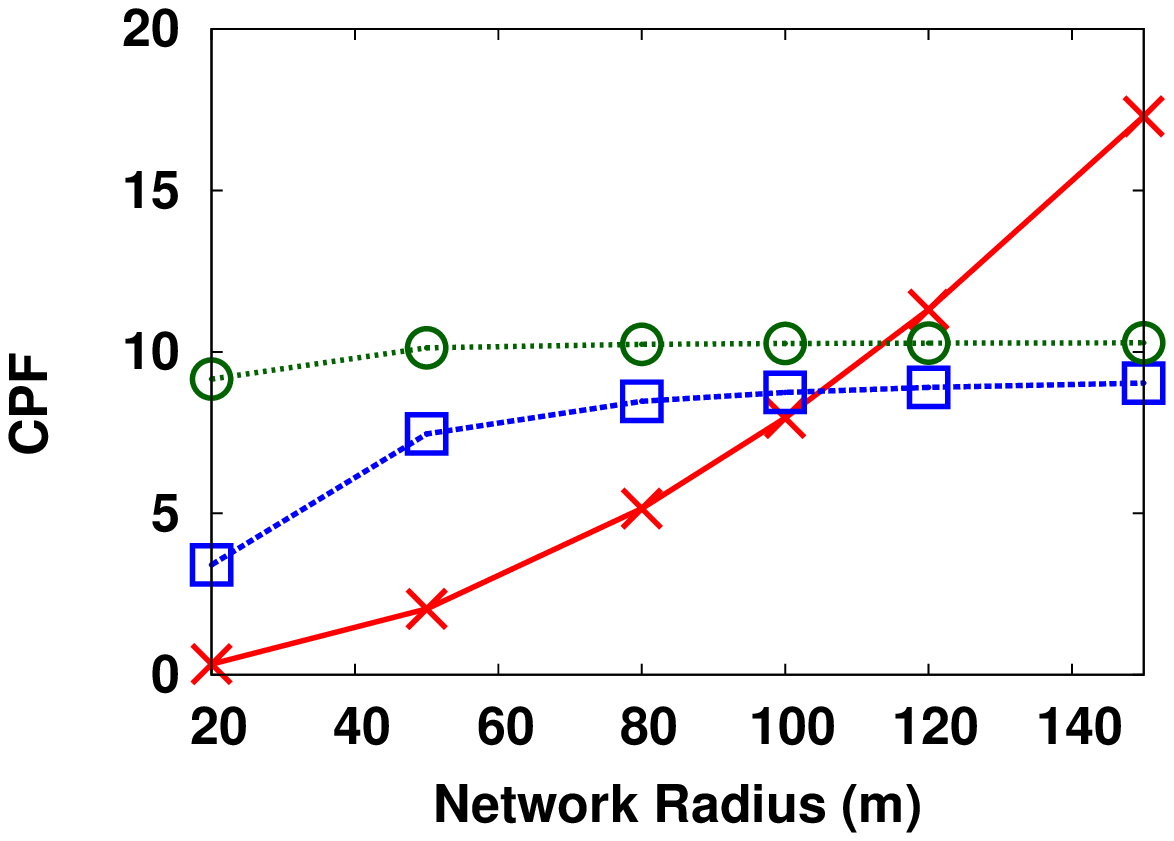}
                \caption{\tiny{ }}
                \label{fig:TSMPER}
        \end{subfigure}
    \vspace{-5mm}
    \caption{\small{CPF ($\alpha=\frac{10}{11}$, $\beta=\frac{1}{11}$) for varying a) packet generation rates (packet/sec), b) node populations, c) network radii (network density).}}
    \label{fig:CPF}
\end{figure}

Next step after deriving the the performance model for each criterion and each category, is computing the $CPF$. 
Figure~\ref{fig:CPF} shows the $CPF$ of the current categories of protocols for $\alpha = \frac{10}{11}$ and $\beta=\frac{1}{11}$ under different conditions. With these values and the $CPF$, we can arrive at the intuitive rules: preamble sampling protocols have a better behavior when the network packet generation rate is low. Scheduled protocols show a better performance when the number of nodes is low in the network. However, their $CPF$ decrease rapidly when the network population increases. For medium packet generation rate, common active period protocols are the best choice.

\begin{table}
\begin{scriptsize}
\begin{center}
\begin{small}
\begin{tabular}{|l||c|c|c|}
 \hline
 &\algname{ScP}&\algname{CAP}&\algname{PSP} \\ \hline
 Case 1&9.22&7.47&6.68 \\ \hline
 Case 2&3.22&5.16&5.92 \\ \hline
\end{tabular}
\end{small}
\end{center}
\end{scriptsize}
\vspace{-3mm}
\caption{CPF comparison between three aforementioned scenarios in Example~\ref{example:1}.} 
\label{table:case}
\end{table}

Let us now revisit Example~\ref{example:1} (mentioned in Section~\ref{sec:intro}).
We use the protocol pool and information presented in Table~\ref{table:1}.
Here the requirements, $R$, are "over hearing avoidance" and having a "distributed" manner. Following Algorithm~\ref{alg:basic}, $\Psi=\{$\algname{ScP},\algname{PSP}$\}$.
Given that energy-consumption is the main concern in this example, we considered  the values $\alpha = \frac{10}{11}$ and $\beta=\frac{1}{11}$ (Please note that $\alpha = \rho_E \times \kappa_E$ and $\beta = \rho_{T_\delta} \times \kappa_{T_\delta}$ are the combination of cost and importance functions--CR: Section~\ref{sec:cpf}).  Table~\ref{table:case} presents the $CPF$ of these categories for the two scenarios of the example (we also added a column for \algname{CAP}). As shown in Table~\ref{table:case}, \algname{ScP} is better for Scenario 1 and \algname{PSP} for Scenario 2. Finally, based on the requirements, $R$, and considering Table~\ref{table:1}, \emph{SMACs} and \emph{AS-MAC} are selected for Scenario 1 and \emph{STEM} for Scenario 2.

To make our $CPF$ model available to WSN designers, we have created an online calculator that can be used to determine performance characteristics of MAC protocols. This tool can be found at the hyper-link of \cite{asudeh}.
\section{Simulation Study}
In order to verify the mathematical energy and delay models derived in the previous sections, we devised a simulation study using a discrete event simulator. This way we are be able to access and modify the underlying parameters of protocols. We compare the results obtained from simulation experiments to the values predicted by our analytical model.  Each data point represents an average of multiple runs; for each data point enough simulations are run to claim at least $95\%$ confidence that the relative error is less than $5\%$. Nodes are randomly deployed in a $100m\times 100m$ area, each with a $20m$ transmission range. In order to reduce the simulation burden, we have used a custom built C++ discrete event simulator. We acknowledge that there are simulation packages that model WSNs, however each of these simulation packages serve a general purpose and have their own idiosyncrasies to overcome. As our goal here was to validate our mathematical models, we elected to program our own simulations that way ensuring that only relevant parts and to the required detail are modeled.

\begin{figure}
        \centering
        \begin{subfigure}[b]{0.33\textwidth}
                \includegraphics[width=\textwidth]{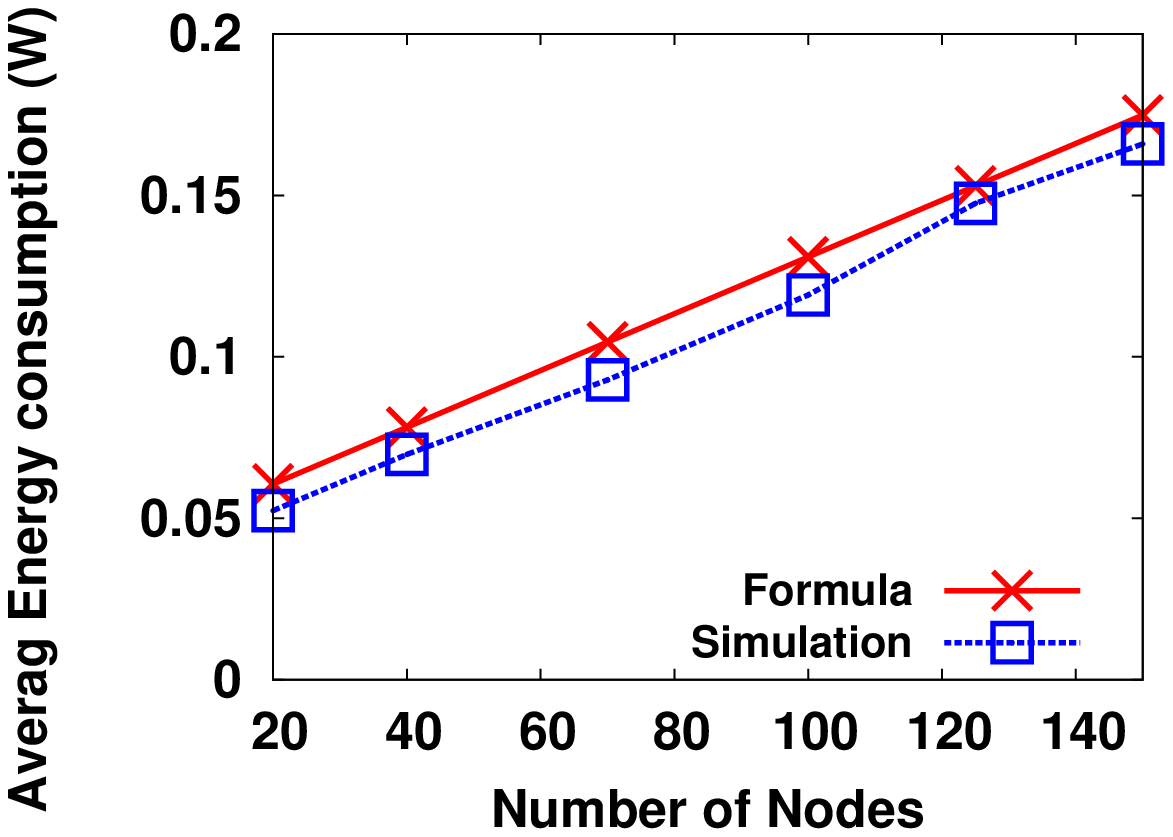}
                \caption{\small{PSA - Energy}}
                \label{fig:PSASIME}
        \end{subfigure}%
        ~
        \begin{subfigure}[b]{0.33\textwidth}
                \includegraphics[width=\textwidth]{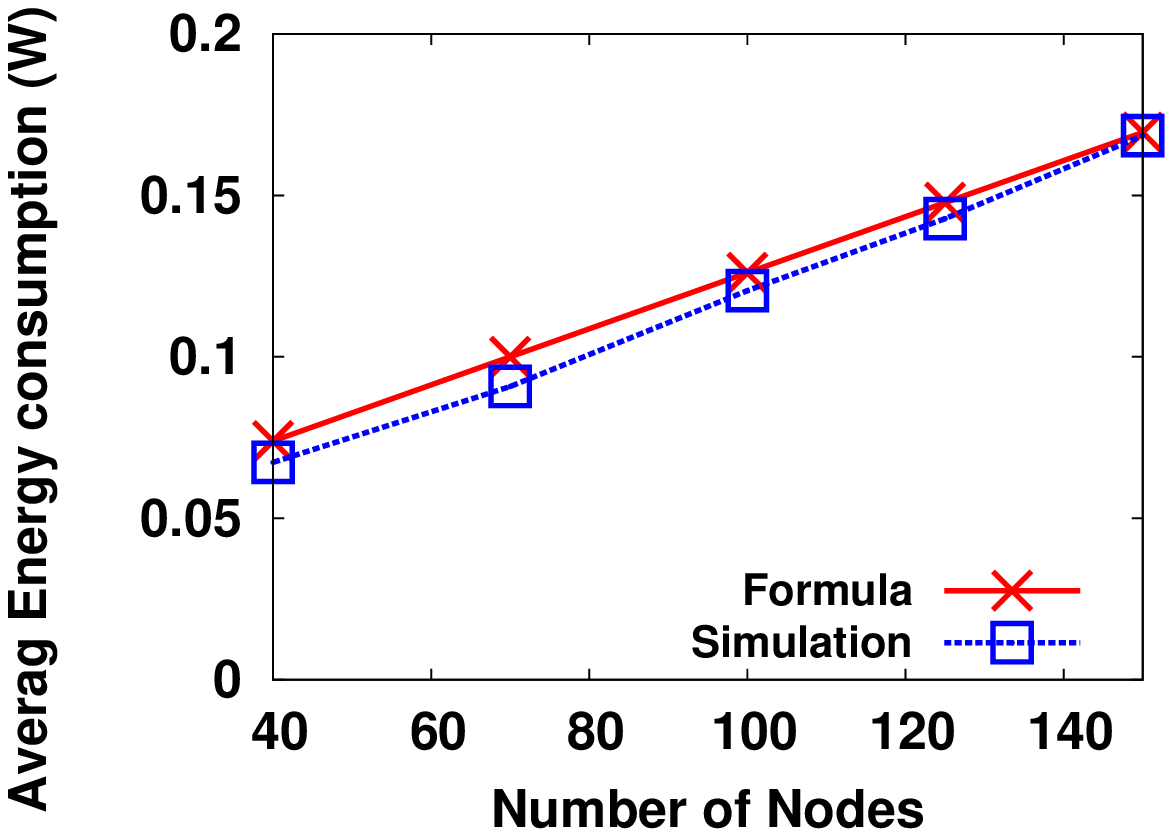}
                \caption{\small{SMAC - Energy}}
                \label{fig:SMACSIME}
        \end{subfigure}%
        ~
        \begin{subfigure}[b]{0.33\textwidth}
                \includegraphics[width=\textwidth]{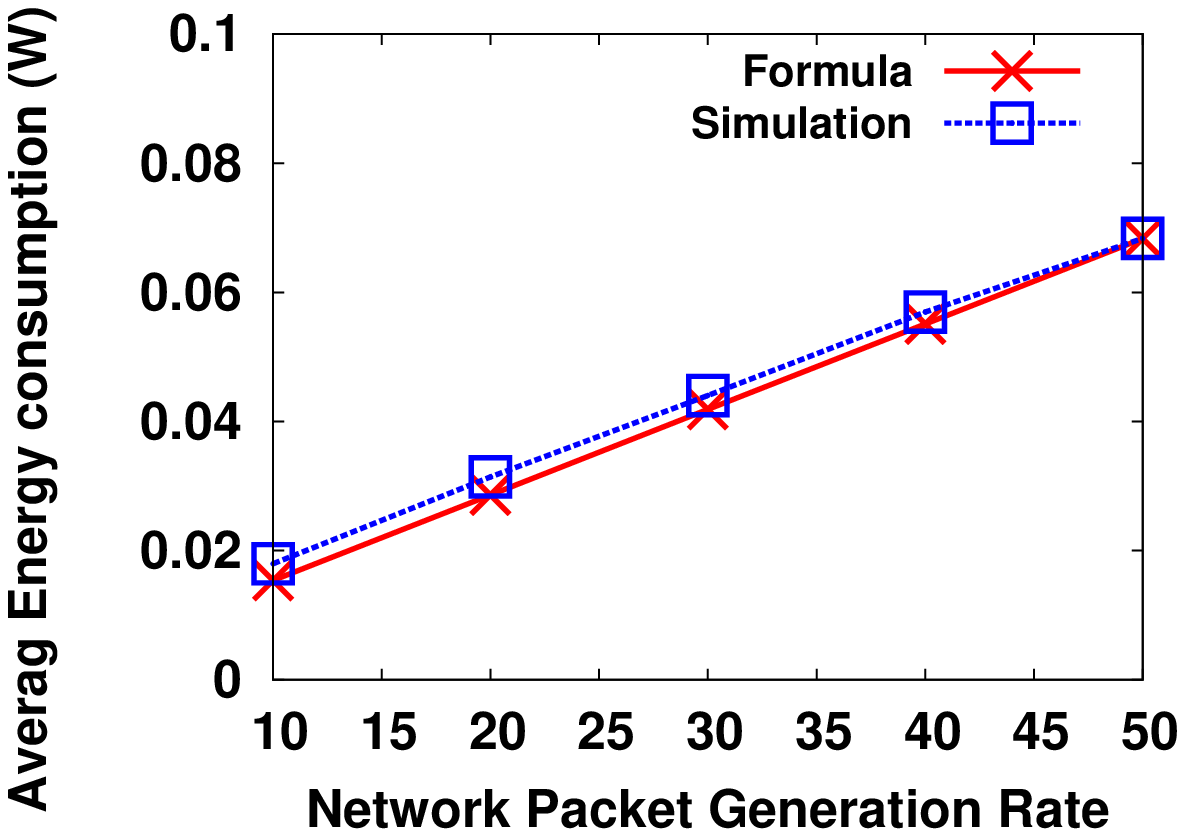}
                \caption{\small{TSMP - Energy}}
                \label{fig:TSMPSIME}
        \end{subfigure}
        ~
        
        \begin{subfigure}[b]{0.33\textwidth}
                \includegraphics[width=\textwidth]{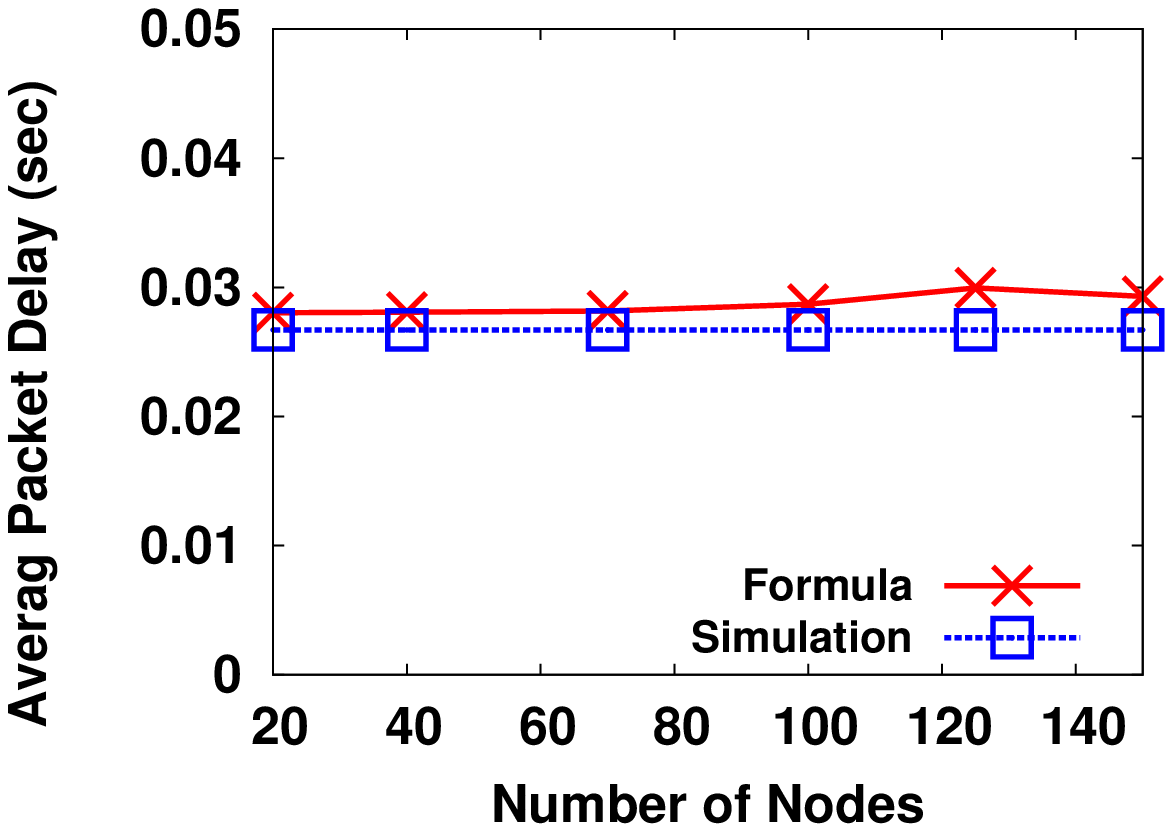}
                \caption{\small{PSA - Delay}}
                \label{fig:PSASIMD}
        \end{subfigure}%
        ~
        \begin{subfigure}[b]{0.33\textwidth}
                \includegraphics[width=\textwidth]{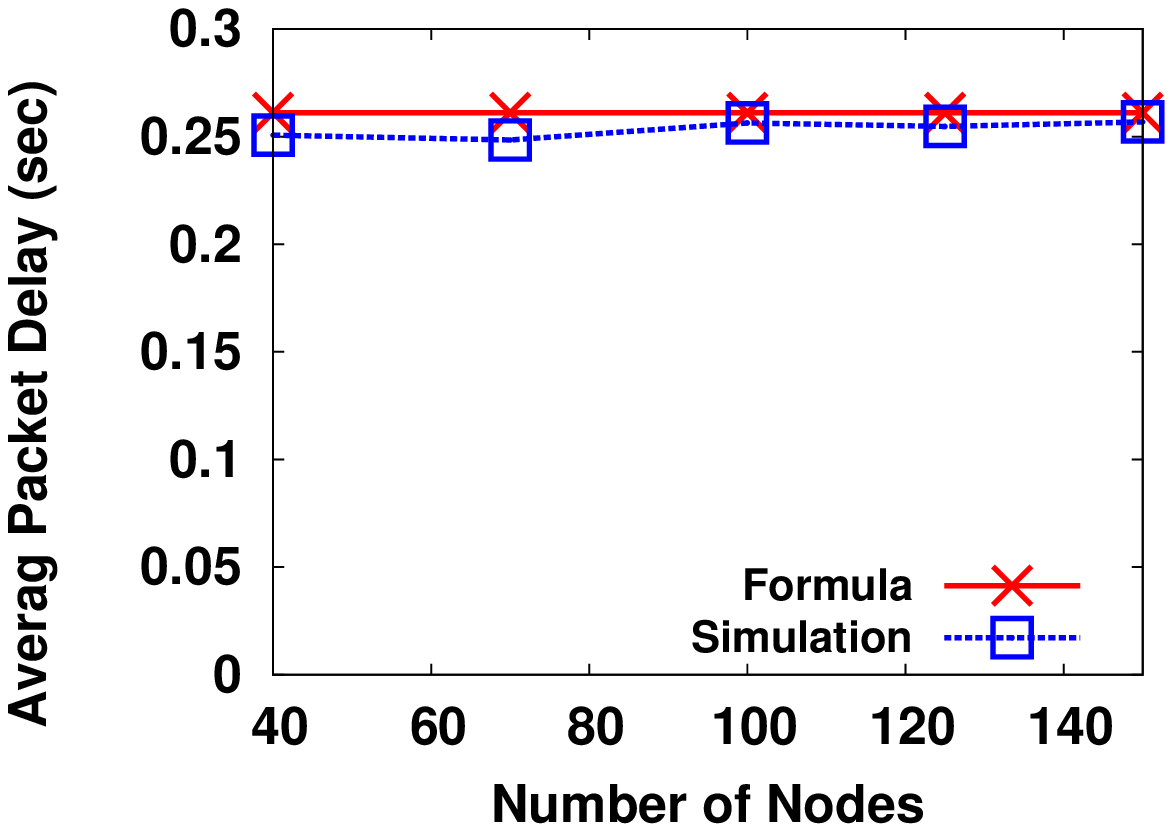}
                \caption{\small{SMAC - Delay}}
                \label{fig:SMACSIMD}
        \end{subfigure}%
        ~
        \begin{subfigure}[b]{0.33\textwidth}
                \includegraphics[width=\textwidth]{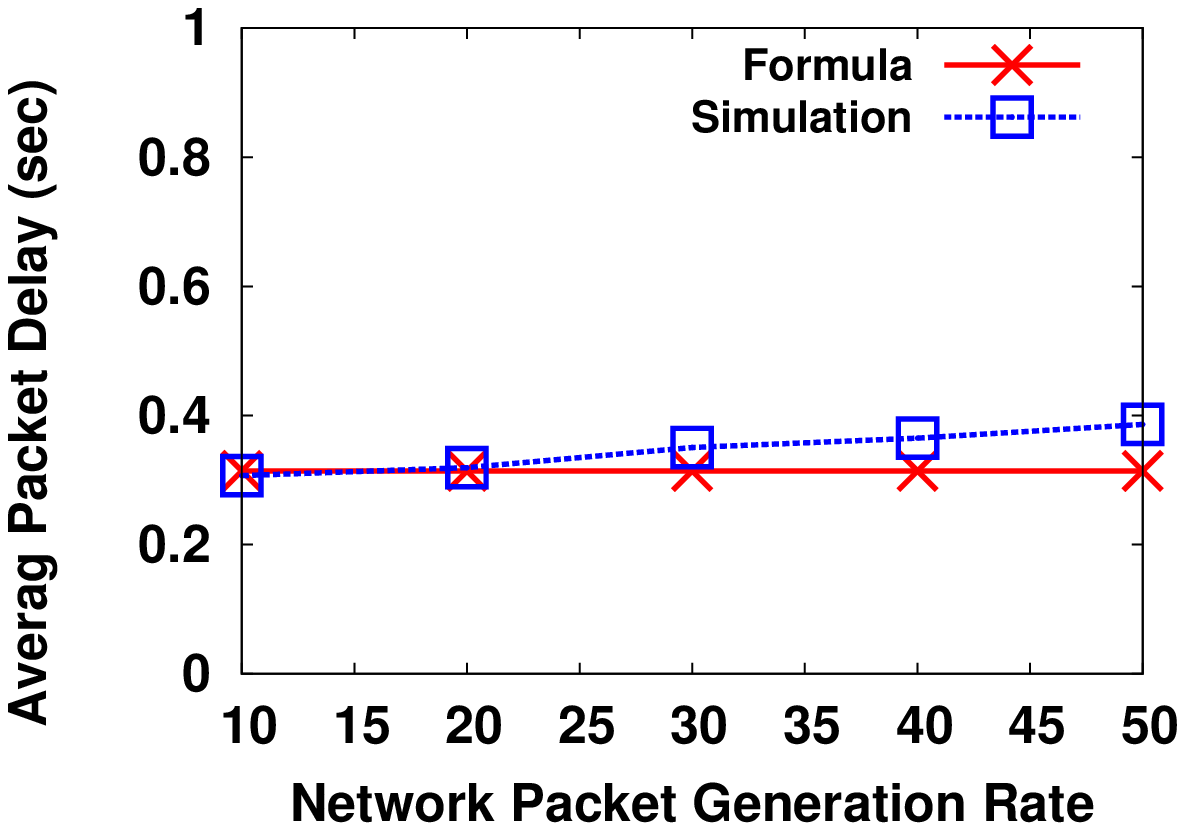}
                \caption{\small{TSMP - Delay}}
                \label{fig:TSMPSIMD}
        \end{subfigure}
    \vspace{-5mm}
    \caption{\small{Model prediction and simulation result comparison with regard to Energy Consumption (first row) and Delay (second row).}}
    \label{fig:SIM}
\end{figure}

The packet generation follows a Poisson distribution with a rate of $\lambda = 20$ packets per second with an available channel bandwidth of $B = 256kbps$. We use the same energy and other parameters (except the network density) we have in Table~\ref{table:notations}. Figure~\ref{fig:SIM} shows the simulation results versus model prediction for the representative protocols for \algname{CAP}, \algname{PSP}, and \algname{ScP}. The first row shows the plots for energy consumption and the second row presents the plots for delay.

The simulation results in Figure~\ref{fig:PSASIME} diverge less than $6.74\%$, validating model predictions for PSA energy consumption. Figure~\ref{fig:PSASIMD} compares the predicted average packet delivery delay in PSA between simulation results and the delay model. Although delay due to queuing is not considered in our model, the results are close to each other. This is because the queues of the nodes are insignificant at low loads.

Figures~\ref{fig:SMACSIME} and \ref{fig:SMACSIMD} show the results from simulation versus model prediction for SMAC, the representative common active period protocol. We focused on the steady state, assuming that nodes already have agreed on a schedule.  Figure~\ref{fig:SMACSIME} shows the average energy consumption per second obtained by simulation and our model. The maximum difference in the two models is $5.42\%$. Figure~\ref{fig:SMACSIMD} presents results for average packet delivery delay obtained from both simulations and analytical model. Again, since the node queues are mostly empty most of the time, the simulation results validate the approximate delay model.

Note that TSMP uses centralized prescheduling; so we created a schedule for a network which contains $10$ connected nodes. The super frame is a table of $3$ rows (frequency division) and $30$ columns (time division). The length of super frame is $0.58875\, sec$, nodes are randomly deployed in a $14m\times 14m$ area, and the transmission range is $20m$ (other parameters are as listed in Table~\ref{table:notations}).
Figure~\ref{fig:TSMPSIME} shows the simulation results for average energy consumption per second in scheduled protocols as well as the results of energy consumption model. The simulation results validate our derived energy consumption model.
Figure~\ref{fig:TSMPSIMD} presents a comparison between the average packet delivery delay from simulation results and model.
\section{Conclusion}
Wireless sensor networks in general are used to sense the broadly defined environment and relay/store such sensed information for processing. This means that WSN application scenarios can be vastly different within diverse environments and requirements. The designers of a WSN need to spend considerable time to decide which MAC protocol(s) to employ as MAC protocols play a paramount role on the network performance. In general, the MAC protocol to be used is selected by rules of thumbs depending on the WSN requirements and scenarios. We argue that such rules of thumb are not sufficient to arrive at the best applicable MAC protocol and parameters to be used for this MAC protocol. We acknowledge that having precise models for all proposed MAC protocols for WSN would be a daunting task. Our goal is to provide a decision-making tool that can help designers select the best MAC protocol and parameters based on some categorization of MAC protocols. In this paper we provided a generic model for selecting MAC protocols for WSNs. The model selects the protocols that satisfy the requirements from the category of protocols that perform better in a given context. We defined the Combined Performance Function to determine the performance of MAC categories under different application scenarios. We also discussed the model expandability over adding new protocols, categories, requirements, and performance criteria. Considering the energy consumption and delay as the initial performance criteria we derived the performance model for three protocol categories of the model. The de-facto rules of thumb of MAC selection closely match our model. We have also created an online version of the CPF model for being able to select the MAC protocols for different contexts \citep{asudeh}.

In this paper we mainly focused on deriving performance models. However preparing the table of protocols and requirements is another important part that has to be done. Moreover, gradually extending the current model using new requirements, protocols, categories, and performance criteria is among the future works. We defined the CPF for average packet generation rate, however, analyzing the behavior of representative MAC categories under different packet assumptions distributions would be more realistic. This is among our future work.

\small
\bibliographystyle{elsarticle-num}
\bibliography{reference-MACSelection}

\end{document}